\documentclass[twocolumn,showpacs,preprintnumbers,nofootinbib,prd,
superscriptaddress, amsmath,10pt]{revtex4-1}
\usepackage{epsf,graphicx,latexsym,multirow,rotating}
\usepackage[usenames,dvipsnames]{xcolor}
\usepackage{hyperref}
\usepackage{aas_macros}
\usepackage[normalem]{ulem}

\newcommand{\dd}{\ensuremath{\mathrm{d}}}
\newcommand{\diff}[2]{\ensuremath{\frac{\dd {#1}}{\dd {#2}}}}
\newcommand{\msun}{\ensuremath{M_\odot}}

\begin{document}

\title{Science with the space-based interferometer LISA. V. Extreme
  mass-ratio inspirals}

\author{Stanislav~Babak}
\affiliation{Max Planck Institut fuer Gravitationsphysik, Albert-Einstein-Institut Am Muehlenberg 1,  D-14476 Golm, Germany}
\author{Jonathan Gair}
\affiliation{School of Mathematics, University of Edinburgh, The King's Buildings, Peter Guthrie Tait Road, Edinburgh, EH9 3FD, UK}
\author{Alberto~Sesana}
\affiliation{School of Physics and Astronomy, University of Birmingham, Edgbaston, Birmingham B15 2TT, UK}
\author{Enrico~Barausse}
\affiliation{Institut d'Astrophysique de Paris, Sorbonne Universit\'es, UPMC Universit\'e Paris 6 \& CNRS, UMR 7095, 98 bis Bd Arago, 
75014 Paris, France}
\author{Carlos~F.~Sopuerta}
\affiliation{Institut de Ci\`encies de l'Espai (CSIC-IEEC), Campus UAB,
Carrer de Can Magrans s/n, 08193 Cerdanyola del Vall\`es, Spain}
\author{Christopher~P.~L.~Berry}
\affiliation{School of Physics and Astronomy, University of Birmingham, Edgbaston, Birmingham B15 2TT, UK}
\author{Emanuele~Berti}
\affiliation{Department of Physics and Astronomy, The University of 
Mississippi, University, MS 38677, USA}
\affiliation{CENTRA, Departamento de F\'isica, Instituto Superior
T\'ecnico, Universidade de Lisboa, Avenida Rovisco Pais 1,
1049 Lisboa, Portugal}
\author{Pau~Amaro-Seoane}
\affiliation{Institut de Ci\`encies de l'Espai (CSIC-IEEC), Campus UAB,
Carrer de Can Magrans s/n, 08193 Cerdanyola del Vall\`es, Spain}
\affiliation{Institute of Applied Mathematics, Academy of Mathematics and Systems Science, CAS, Beijing 100190, China\\
Kavli Institute for Astronomy and Astrophysics, Beijing 100871, China\\
Zentrum f{\"u}r Astronomie und Astrophysik, TU Berlin, Hardenbergstra{\ss}e 36, 10623 Berlin, Germany}
\author{Antoine Petiteau}
\affiliation{APC, Universit\'e Paris Diderot, Observatoire de Paris, Sorbonne
Paris Cit\'e,
10 rue Alice Domon et L\'eonie Duquet, 75205 Paris Cedex 13, France}
\author{Antoine Klein}
\affiliation{Institut d'Astrophysique de Paris, Sorbonne Universit\'es, UPMC Universit\'e Paris 6 \& CNRS, UMR 7095, 98 bis Bd Arago, 
75014 Paris, France}

\date{\today}

\begin{abstract}
The space-based Laser Interferometer Space Antenna (LISA) 
will be able to observe the gravitational-wave signals from systems comprised of a
massive black hole and a stellar-mass compact object. These systems are known as
extreme-mass-ratio inspirals (EMRIs) and are expected to complete $\sim 10^4$--$10^5$
cycles in band, thus allowing exquisite measurements of their parameters.
In this work, we attempt to quantify the astrophysical uncertainties affecting the 
predictions for the number of EMRIs detectable by LISA, and find that 
competing astrophysical assumptions produce a variance of about three orders of magnitude in the expected intrinsic EMRI rate. 
However, we find that irrespective of the astrophysical model, at least a few EMRIs per year should be detectable by the LISA mission, with up to a few thousands per year under the most optimistic astrophysical assumptions. 
We also investigate the precision with which LISA will be able to extract the parameters of these sources. We find that
typical fractional statistical errors with which the intrinsic parameters (redshifted masses, massive black hole spin and orbital eccentricity) can be recovered are $\sim 10^{-6}$--$10^{-4}$. 
Luminosity distance (which is required to infer true masses) is inferred to about $10\%$ precision and sky position is localized to a few square degrees, while tests of the multipolar structure of the Kerr metric can be performed to percent-level precision or better. 

\end{abstract}

\maketitle

\section{Introduction}

Gravitational waves (GWs) provide a means of gathering precious
information otherwise beyond the reach of traditional electromagnetic
astronomy.  In particular, GWs may illuminate our understanding of the
properties of black holes (BHs). The terrestrial Advanced
LIGO~\cite{2015CQGra..32g4001L} has recently observed GW signals from
coalescing stellar-mass binary BHs, with two clear
detections~\cite{2016PhRvL.116f1102A,2016PhRvL.116x1103A} and a
probable third
candidate~\cite{2016PhRvD..93l2003A,2016PhRvX...6d1015A}. These
observations allowed estimation of the source parameters with
high
accuracy~\cite{2016PhRvL.116x1102A,2016PhRvX...6d1014A,2016PhRvX...6d1015A},
giving new insight into their astrophysical
formation~\cite{2016ApJ...818L..22A,2016PhRvX...6d1015A} and allowing
tests of general relativity
(GR)~\cite{2016PhRvL.116v1101A,2016PhRvD..94h4002Y,2016PhRvX...6d1015A}. Many
more stellar-mass BH binaries are expected to be detected by LIGO (and by other
terrestrial detectors such as Advanced
Virgo~\cite{2015CQGra..32b4001A} and KAGRA~\cite{2012CQGra..29l4007S})
in the next few years~\cite{2016PhRvX...6d1015A,2016LRR....19....1A}.

In addition to stellar-mass BHs, there is believed to be a population 
of massive BHs (MBHs), with masses in the range
$10^5$--$10^9\,M_{\odot}$, 
each lurking at the center of a galaxy~\cite{1971MNRAS.152..461L,1982MNRAS.200..115S,2010RvMP...82.3121G,1995ARA&A..33..581K,2009ApJ...698..198G}. 
Correlations between the mass of the
MBH and other characteristics of the surrounding stars, such as the
velocity dispersion $\sigma$ of the spheroidal component of the host
galaxy (see, e.g.,~\cite{2013ARA&A..51..511K}) suggest a link between evolution 
of the MBH and its host galaxy~\cite{2003ApJ...596L..27K,2004ASSL..308..147H,2009MNRAS.400.1911V}. 

Surrounding MBHs out to distances of a few parsecs, are nuclear star clusters of 
millions of stars~\cite{2014CQGra..31x4007S}. In these innermost galactic regions, the
density of stars easily exceeds $10^6\,M_{\odot}\,\mathrm{pc}^{-3}$,
and relative stellar velocities range between $\sim$ $100$--$1000~\mathrm{km\,s^{-1}}$. 
Here, mutual gravitational deflections between stars
play a crucial role in determining dynamics~\cite{2005PhR...419...65A}, and their tidal disruption may contribute to
increasing the mass of the central MBH~\cite{1975Natur.254..295H,1991ApJ...370...60M,2002A&A...394..345F,2003ApJ...592...42G}. 
Unlike stars, compact
objects (COs; including stellar-mass BHs, neutron stars and white dwarfs)
can avoid tidal disruption and approach the central MBH, radiating a
significant amount of energy in GWs at low frequencies.

One of the main experimental challenges for ground-based detectors is
seismic noise, which limits their sensitivity at frequencies
$\lesssim 10~\mathrm{Hz}$, making them insensitive to GWs from MBH
systems.  However, space-borne interferometric GW detectors, such as
the Laser Interferometer Space Antenna
(LISA)~\cite{2017arXiv170200786A}, are free from the seismic
noise. The technology behind LISA, an ESA-led mission expected to be
launched by 2034, has been recently tested by the LISA
Pathfinder experiment with outstanding
results~\cite{2016PhRvL.116w1101A}. Previous work has investigated the
scientific potential of LISA-like detectors for (i) MBH mergers and
astrophysics~\cite{2016PhRvD..93b4003K}; (ii) stochastic
backgrounds~\cite{2016JCAP...04..001C,2016JCAP...12..026B}; (iii)
cosmography~\cite{2016JCAP...04..002T}; (iv) tests of general
relativity~\cite{2016PhRvL.116x1104B,2016PhRvL.117j1102B}; and (v)
ringdown tests of the nature of
BHs~\cite{2006PhRvD..73f4030B,2016PhRvL.117j1102B}.  LISA will also
usher in the era of multiband {GW} astronomy, with stellar-mass binary
{BH}s being detectable by {LISA} years to days before they reach the
sensitivity window of ground-based
detectors~\cite{2016PhRvL.116w1102S}. This would provide information
on the formation mechanism of {BH}
binaries~\cite{2016PhRvD..94f4020N,2017MNRAS.465.4375N,2016ApJ...830L..18B},
improve the precision of parameter estimation (including sky
location)~\cite{2016PhRvL.117e1102V}, and yield better constraints on
putative deviations from GR~\cite{2016PhRvL.116x1104B}.  In this paper
we will focus on the physics and astrophysics of \emph{extreme
  mass-ratio inspirals} (EMRIs)~\cite{2007CQGra..24R.113A}, i.e.\
systems comprised of stellar-mass BHs or other comparable mass COs
orbiting around a MBH with mass $M\sim 10^4$--$10^7 \msun$.

As a consequence of their extreme mass ratio these systems inspiral
slowly, completing $\sim 10^4$--$10^5$ cycles in LISA's sensitive
frequency
range~\cite{1964PhRv..136.1224P,2008PhRvD..78f4028H}. Therefore
{EMRI}s are ideal signals to construct detailed maps of the background
spacetime of MBHs~\cite{1995PhRvD..52.5707R,1997PhRvD..56.1845R,1997PhRvD..56.7732R,2006CQGra..23.4167G,2007PhRvD..75d2003B}, precisely estimate source
parameters~\cite{2004PhRvD..69h2005B,2009PhRvD..79h4021H,2009CQGra..26i4027A},
perform tests of GR~\cite{2007PhRvD..75d2003B,2013LRR....16....7G},
and possibly detect the presence of gas around the central
MBH~\cite{2007PhRvD..75f4026B,2008PhRvD..77j4027B,2011PhRvD..83d4037G,2011PhRvL.107q1103Y,2014PhRvD..89j4059B,2015JPhCS.610a2044B}. Measuring
the properties of a population of {EMRI} signals could additionally
give us information on the mass distribution of
{MBH}s~\cite{2010PhRvD..81j4014G} and their host stellar
environments~\cite{2007CQGra..24R.113A}. 

We examine in detail the scientific potential of {EMRI}
observations with {LISA}, focusing on event rates and on parameter-estimation 
precision.  There have been previous studies computing EMRI
rates~\cite{2004CQGra..21S1595G,2009CQGra..26i4034G,2016PhRvD..94l4042B}, but the
astrophysical model employed in those calculations was a combination
of simple power laws, and no attempt was made to quantify the
uncertainties in that model. EMRI parameter-estimation studies have
also been carried out~\cite{2004PhRvD..69h2005B,2009PhRvD..79h4021H},
but only for a small sample of representative cases and not for a full
astrophysical population. In this study we address both of these
shortcomings. We compute event rates for several different
astrophysical models that were selected to quantify the main
observational uncertainties, and we compute estimates of the
parameter-estimation precisions for all the events in each
population. Our results are
computed for the first time considering a $2.5~\mathrm{Gm}$ LISA
detector with six laser links, which was proposed as the new mission
baseline in the response to the ESA call in January 2017~\cite{2017arXiv170200786A}.

The plan of the paper is as follows. We begin in
Section~\ref{sec:sensitivity} by discussing the assumed design of the
LISA detector. In Section~\ref{sec:model} we describe our
astrophysical {EMRI} model and the related
uncertainties. Section~\ref{sec:analysis} describes our {EMRI}
waveform models and the parameter estimation calculation. We summarize
our main results in Section~\ref{sec:results}, and conclude by
presenting possible directions for future research.

\section{LISA sensitivity}
\label{sec:sensitivity}

The LISA baseline went through several stages of re-design in the past
five years. Following the 2011 NASA drop-out, the classic LISA design
was initially descoped to fit within the budget of an L-class ESA
mission, leading to the New Gravitational-wave Observatory (NGO)
design~\cite{2012CQGra..29l4016A}.  This new baseline was eventually
selected as strawman mission in support of \textit{The Gravitational
 Universe}~\cite{2013arXiv1305.5720C}, the science theme adopted by
ESA for its L3 slot, scheduled for launch in 2034.  Following the
selection in 2014, a Gravitational Observatory Advisory Team (GOAT)
was appointed by ESA to consider a number of feasible options and
issue a recommendation for a new design. The study considered a family
of designs, featuring different choices for the arm length $L$, laser
power, telescope diameter, mission duration and low-frequency noise
level (see~\cite{2016PhRvD..93b4003K} for details).

Following the GOAT recommendation, the LISA Consortium answered the
ESA call for missions by proposing the baseline outlined in
\cite{2017arXiv170200786A}. The detector features a constellation of
three satellites separated by $L=2.5~\mathrm{Gm}$ and
connected by six laser links. The output power of each laser is
$2~\mathrm{W}$ and their light is collected by $30~\mathrm{cm}$
telescopes. The sky-averaged detector sensitivity can be written in
analytic form as
\begin{eqnarray}
  S_n(f)&=&\frac{20}{3}\frac{4S_{n}^\mathrm{acc}(f)+2S_{n}^\mathrm{loc}+S_{n}^\mathrm{sn}+S_{n}^\mathrm{omn}}{L^2} \nonumber \\
& & \times 
\left[1+\left(\frac{2Lf}{0.41 c}\right)^2\right], 
  \label{eq:sens}
\end{eqnarray}
where $L$ is the arm length, and the noise contributions
$S_{n}^\mathrm{acc}(f)$, $S_{n}^\mathrm{loc}$,
$S_{n}^\mathrm{sn}$ and $S_{n}^\mathrm{omn}$ are due to
low-frequency acceleration, local interferometer noise, 
shot noise and other measurement noise, respectively. The acceleration
noise has been fitted to the level successfully demonstrated by the
LISA Pathfinder~\cite{2016PhRvL.116w1101A} as
\begin{eqnarray}
  S_{n}^\mathrm{acc}(f) & = & \left\{9 \times 10^{-30}+3.24 \times 10^{-28}\left[\left(\frac{3\times10^{-5}~\mathrm{Hz}}{f}\right)^{10} \right. \right. \nonumber \\
  &  & \left. \left. + \left(\frac{10^{-4}~\mathrm{Hz}}{f}\right)^{2}\right]\right\}\frac{1}{(2\pi f)^4}\,\mathrm{{m^2\,Hz}^{-1}},
\end{eqnarray}
whereas other contributions are set to
\begin{equation}
\begin{split}
&S_{n}^\mathrm{loc}= 2.89\times10^{-24}~\mathrm{{m}^2\,{Hz}^{-1}},\\
&S_{n}^\mathrm{sn}= 7.92\times10^{-23}~\mathrm{{m}^2\,{Hz}^{-1}},\\
&S_{n}^\mathrm{omn}=4.00\times10^{-24}~\mathrm{{m}^2\,{Hz}^{-1}}.
\end{split}
\end{equation}

Besides the instrumental noise of Eq.~(\ref{eq:sens}), we also include
a galactic confusion noise component, modeled by the fit
\begin{eqnarray}
    S_\mathrm{gal} &=& A_\mathrm{gal} \left(\frac{f}{1~\mathrm{Hz}}\right)^{-7/3} \exp\left[-\left( \frac{f}{s_1}\right)^{\alpha}\right] \nonumber \\
    & & \times \frac1{2} \left[1+\tanh\left(-\frac{f-f_0}{s_2}\right)\right]\,.
\end{eqnarray}
The overall amplitude of the background
$A_\mathrm{gal}=3.266\times 10^{-44}~\mathrm{{Hz}^{-1}}$ depends on
the astrophysical model for the population of white dwarf binaries in
the Galaxy. Here we have used the same model as in
\cite{2017arXiv170200786A}. The power law $f^{-7/3}$ is what we
expect from a population of almost monochromatic binaries. The
exponential factor comes from
removal of the loud signals standing above the confusion background,
while the last term takes
into account that all Galactic binaries can be resolved and
removed above some frequency $f_0$. For the assumed two-year
observation period, the fitting parameters appearing in the above
expression for $S_\mathrm{gal}$ have the values:
$\alpha = 1.183$, 
$s_1 = 1.426~\mathrm{mHz}$, 
$f_0 = 2.412~\mathrm{mHz} $, 
$s_2 = 4.835~\mathrm{mHz}$. 

The LISA design is most sensitive at millihertz frequencies, making it well-purposed for observing EMRIs.


\section{Astrophysical EMRI model}
\label{sec:model}

The expected EMRI rate depends on several astrophysical ingredients:
\begin{itemize}
\item The MBH population in the accessible LISA mass range, $M \in \left[10^4, 10^7\right] \msun$, the redshift evolution of their mass function, and their spin distribution;
\item The fraction of MBHs hosted in dense stellar cusps, which are the nurseries for EMRI formation;
\item The EMRI rate per individual MBH, and the mass and eccentricity distribution of the inspiralling COs.
\end{itemize}
In the following subsections we consider these ingredients in turn, presenting the astrophysically motivated prescriptions used in this work, before combining them in Section~\ref{sec:bakery}.
  
\subsection{MBH population}
\label{sec:mbhpop}
\begin{figure}
\centering
\includegraphics[width=\columnwidth,clip=true,angle=0]{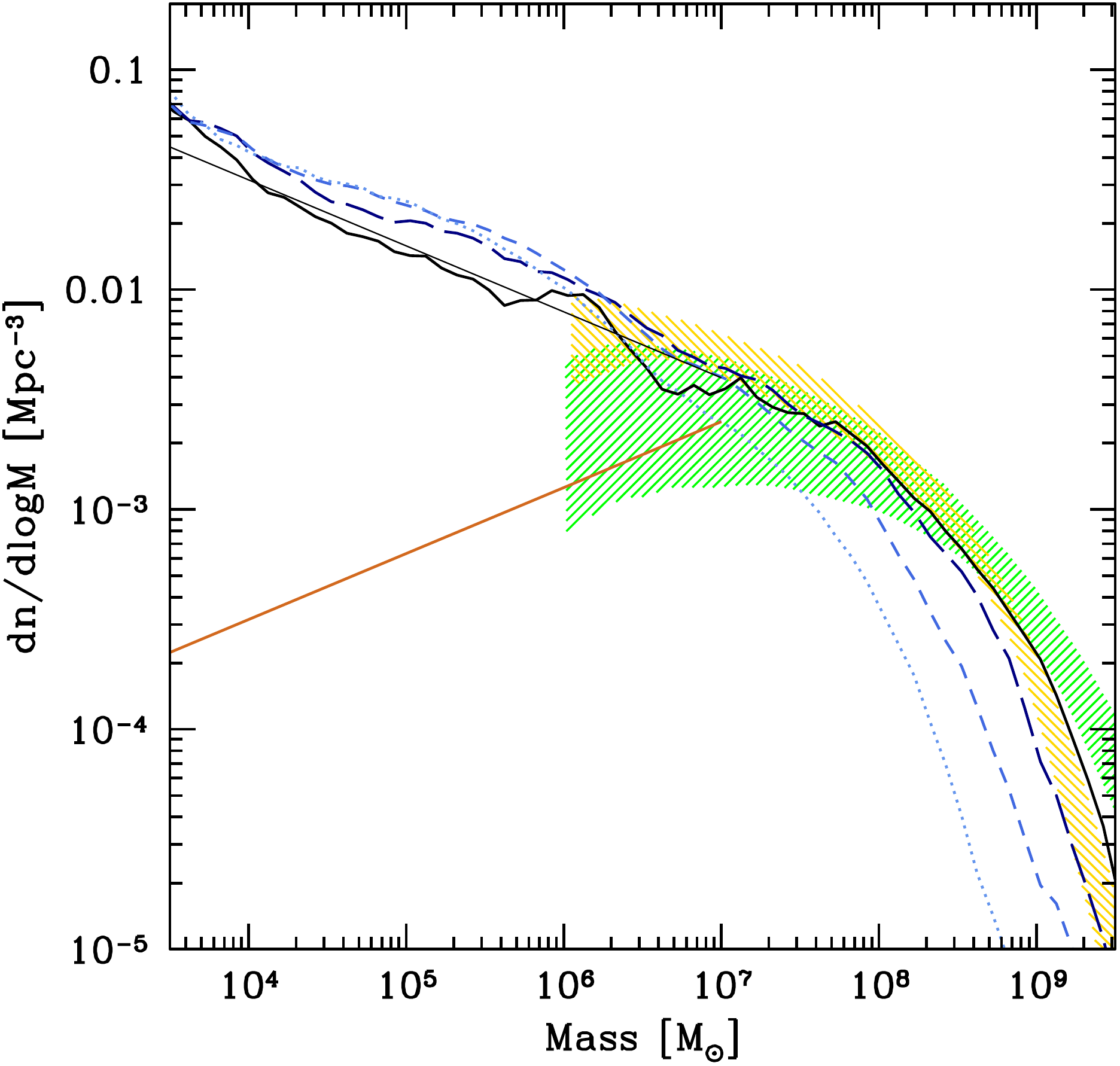}
\caption{MBH density mass function $\dd n/\dd\, \log_{10}M$ for the self-consistent model popIII at redshift $0$ (solid), $1$ (long dashed), $2$ (short dashed) and $3$ (dotted). 
The approximation provided by Eq.~(\ref{mfapprox}) is shown as a thin straight black line. Also shown in brown is the redshift-independent pessimistic 
mass function as given by Eq.~(\ref{mfpess}). The shaded area represent constraints from Shankar et al.~\cite{2009ApJ...690...20S} (light orange) and Shankar~\cite{2013CQGra..30x4001S} (green).}
\label{mfuncnew}
\end{figure}

We consider here two population models that are intended to bracket current uncertainties in the MBH mass function at the low mass end (cf. Figure~\ref{mfuncnew}). 
The first one is \emph{Model popIII}, as investigated in Klein et al.~\cite{2016PhRvD..93b4003K}. This is a self-consistent model for MBH formation and 
cosmic evolution developed in~\cite{2012MNRAS.423.2533B,2014ApJ...794..104S,2015ApJ...806L...8A,2015ApJ...812...72A}, and assumes light MBH 
seeds from population III (popIII) stars~\cite{2001ApJ...551L..27M}, while accounting for the delays between MBH and galaxy mergers. The model 
successfully reproduces several galaxy and MBH mass function properties, and it is consistent with observational constraints on the MBH mass 
function~\cite{2009ApJ...690...20S,2013CQGra..30x4001S}. The predicted MBH mass function in the relevant range can be approximated as
\begin{equation}
  \diff{n}{\,\log M} = 0.005 \left( \frac{M}{3\times 10^6 \msun} \right)^{-0.3}~\mathrm{Mpc^{-3}},
  \label{mfapprox}
\end{equation}
almost independent of redshift, as shown in Figure~\ref{mfuncnew}. We label this mass function ``Barausse12''. 

Following Gair et al.~\cite{2010PhRvD..81j4014G}, we also consider a more conservative model with a redshift-independent mass function of the form
\begin{equation}
\diff{n}{\,\log M} = 0.002 \left( \frac{M}{3\times 10^6 \msun} \right)^{0.3}~\mathrm{Mpc^{-3}}.
  \label{mfpess}
\end{equation}
In this case, the MBH mass function \emph{increases} with mass at the low-mass end, and it is therefore less favorable for EMRI events 
falling in the LISA band. This is a purely phenomenological model, which 
does not come from a self-consistent MBH evolutionary scenario, 
but is still consistent with current observational constraints on the MBH mass function. We label this mass function ``Gair10''.

The EMRI rate and expected signal also depend on the spin parameter $a$ of the central MBH. The popIII model self-consistently follows the spin 
evolution of MBHs through accretion and mergers. We find that most MBHs in the LISA range have near maximal spins. This is because, although MBH seeds
start with a random spin distribution, they need to accrete enough mass to get into the LISA band. At small masses, the MBHs always align with the 
accreted material (the MBH angular momentum is always smaller than the disk angular momentum in our model~\cite{2014ApJ...794..104S}). 
The distribution starts to extend to lower $a$ for higher masses,
when the MBH spin becomes larger than the typical disk angular momentum~\cite{2013ApJ...762...68D}. As a result, 
the MBHs do not always align with the accreting material, and spindown is possible. 
However, this effect becomes appreciable only at $M\approx10^7\msun$~\cite{2013ApJ...762...68D,2014ApJ...794..104S}. We assume a maximum dimensionless MBH 
spin parameter $a=0.998$, with a median value around $a=0.98$. Since most MBHs have high spins in our default model, we label it ``a98''. For the sake of comparison, we also consider
two alternative models; one with a flat spin distribution $0<a<1$, labeled ``aflat'', and one with nonspinning MBHs, labeled ``a0''.

\subsection{Stellar cusps surrounding MBHs}
A necessary condition for EMRI formation is the presence of a cusp-like distribution of stellar objects surrounding the MBH. It has generally been assumed 
that MBHs are immersed in a Bahcall--Wolf stellar cusp with density profile $\rho(r)\propto r^{-7/4}$, which is the steady state solution for a distribution 
of stars in the sphere of influence of a massive object~\cite{1977ApJ...216..883B}. However galaxies merge, and so do the MBHs they host. MBH binaries destroy 
stellar cusps, carving a low density core~\cite{2001ApJ...563...34M,2015ApJ...806L...8A,2015ApJ...812...72A} which is unsuitable to the formation of EMRIs. One of the main advantages of using a semi-analytic
MBH evolution model is that we are able to track the MBH merger history implementing a simple prescription that takes into account in a self-consistent way cusp disruption 
following MBH binary mergers.

\subsubsection{Cusp erosion and regrowth.}
\begin{figure*}
\centering
\includegraphics[width=\columnwidth,clip=true,angle=0]{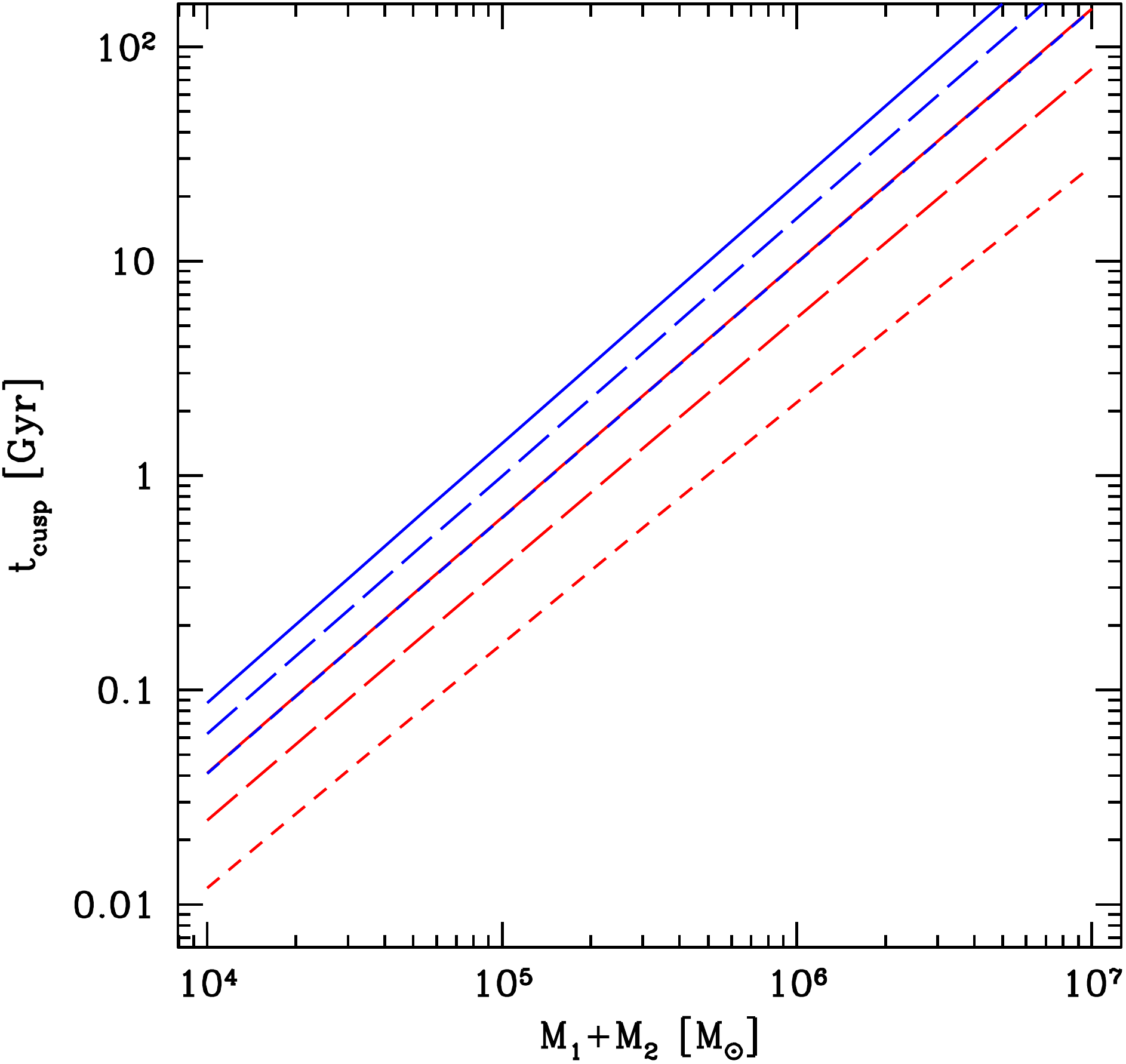} \quad 
\includegraphics[width=\columnwidth,clip=true,angle=0]{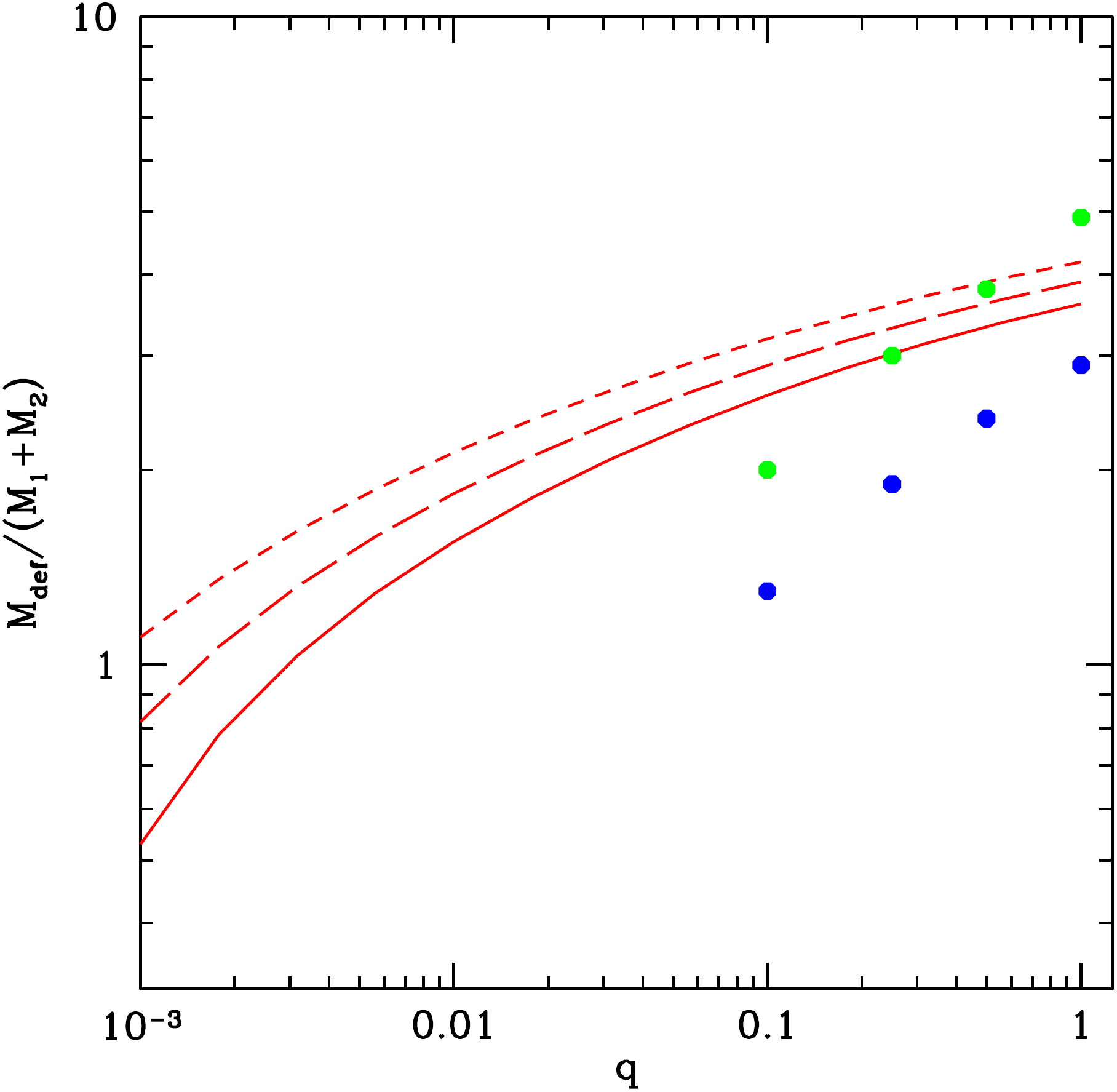}\\
\caption{Left panel: Cusp regrowth time $t_\mathrm{cusp}$ as a function of the total MBH binary mass. Solid, long-dashed and short-dashed 
curves are for $q=1,0.1,0.01$ respectively. Red curves assume $V_\mathrm{k}=0$ whereas blue curves assume $V_\mathrm{k}/V_\mathrm{esc}=0.6$. 
Right panel: Mass deficit normalized to $M$ as a function of binary mass ratio for $M=10^5\msun$(short dashed), $M=10^6\msun$ (long dashed), 
and $M=10^7\msun$ (solid). Blue and green dots are mass deficits computed by Khan et al.~\cite{2012ApJ...749..147K}.}
\label{tcusp}
\end{figure*}
To understand the impact of a merger we must estimate the time $t_\mathrm{cusp}$ taken for a cusp to reform.

We assume that each MBH binary with mass $M=M_1+M_2$ is embedded in an
isothermal sphere, defined by a density profile~\cite{1987gady.book.....B}
\begin{equation}
  \rho(r)=\frac{\sigma^2}{2\pi G r^2},
\end{equation}
where $\sigma$ is the one-dimensional velocity dispersion. We further assume that the MBH binary carves a core of constant density and size $r_\mathrm{c}$ in the 
center of the stellar system. The mass deficit due to a flat core of size $r_\mathrm{c}$ is given by 
\begin{equation}
  M_\mathrm{d}=\frac{4}{3}\frac{\sigma^2r_\mathrm{c}}{G}.
  \label{mrc}
\end{equation}
Thus, $r_\mathrm{c}$ can be estimated once $M_\mathrm{d}$ is known. The mass deficit must equal the mass displaced by the MBH binary on its way to
coalescence, and is estimated as~\cite{2015ApJ...806L...8A,2015ApJ...812...72A}:
\begin{equation}
  M_\mathrm{d}=0.7Mq^{0.2}+0.5M{\ln}\left(\frac{r_\mathrm{h}}{r_\mathrm{GW}}\right)+5M\left(\frac{V_\mathrm{k}}{V_\mathrm{esc}}\right)^{1.75}.
  \label{mdeficit}
\end{equation}
Here, $q = M_2/M_1 \leq 1$ is the mass ratio of the MBH binary, $r_\mathrm{h}$ is the binary hardening radius, $r_\mathrm{GW}$ is the radius at which GW emission dominates over stellar hardening, $V_\mathrm{k}$ 
is the GW kick and $V_\mathrm{esc}\approx5\sigma$ is the typical escape velocity from the stellar bulge~\cite{2006ApJ...651..392S}. To make use of Eq.~(\ref{mdeficit}), we need an estimate of $r_\mathrm{h}/r_\mathrm{GW}$. Here $r_\mathrm{h}$ is the hardening radius, the separation at which the specific binding energy of the binary is equal to the average specific kinetic energy of the surrounding stars~\cite{1975MNRAS.173..729H,1975AJ.....80..809H}, given by
\begin{equation}
r_\mathrm{h}=\frac{GM_2}{4\sigma^2},
\end{equation}
where $M_2$ is the secondary's mass  (see e.g. \cite{1996NewA....1...35Q}).
The distance $r_\mathrm{GW}$ represents the separation at which the MBH binary evolution switches from being stellar hardening dominated to be GW driven. It can therefore be computed by finding where the three-body scattering hardening rate $(\dd r/\dd t)_*$ becomes equal to the GW shrinking rate $(\dd r/\dd t)_\mathrm{GW}$. The latter given by the standard quadrupole formula~\cite{1964PhRv..136.1224P}, and the former can be written as~\cite{1996NewA....1...35Q}
\begin{equation}
\left(\diff{r}{t}\right)_*=\frac{HG\rho_*}{\sigma}r^2.
\end{equation}
Here $H\approx 15$ is a dimensionless hardening rate and the stellar density $\rho_*$ is evaluated at the influence radius of the binary $r_\mathrm{i}=GM/\sigma^2$~\cite{2015MNRAS.454L..66S}. For the isothermal sphere this gives 
\begin{equation}
\rho_*=\frac{\sigma^6}{2\pi G^3 M^2}.
\end{equation}
Combining everything together and assuming circular binaries, one gets:
\begin{equation}
  \frac{r_\mathrm{h}}{r_\mathrm{GW}}\approx 0.178 \frac{c}{\sigma}\frac{q^{4/5}}{(1+q)^{3/5}}.
  \label{rgwsurc}
\end{equation}
For a given MBH mass, if we know that there was a merger with a given $q$, we can substitute Eq.~(\ref{rgwsurc}) into Eq.~(\ref{mdeficit}), 
to obtain the mass deficit, and use this in Eq.~(\ref{mrc}) to solve for $r_\mathrm{c}$ and obtain the extent of the core.
Once $r_\mathrm{c}$ is known, the relaxation time for an isothermal sphere is given by~\cite{1987gady.book.....B}
\begin{equation}
  t_\mathrm{relax}=\frac{5}{\ln\Lambda}\left(\frac{\sigma}{10~\mathrm{km\,s^{-1}}}\right)\left(\frac{r_\mathrm{c}}{1~\mathrm{pc}}\right)^2\mathrm{Gyr},
  \label{trelax}
\end{equation}
where $\ln\Lambda\approx 10$ is the Coulomb logarithm~\cite{1987gady.book.....B}. The cusp regrowth time is then~\cite{2011CQGra..28i4017A}
\begin{equation}
  t_\mathrm{cusp} = 0.25 t_\mathrm{relax}.
  \label{tcuspr}
\end{equation}
This can be expressed in terms of $M$ and $q$ only if we specify an $M$--$\sigma$ relation to eliminate the $\sigma$ dependence. We use the best fit of G{\"u}ltekin et al.~\cite{2009ApJ...698..198G} as our default model:
\begin{equation}
  M=1.53 \times 10^6 \left(\frac{\sigma}{70~\mathrm{km\,s^{-1}}}\right)^{4.24} \msun.
\end{equation}
Combining Eq.~(\ref{mrc})--(\ref{tcuspr}), it is possible to approximate $t_\mathrm{cusp}\propto M^{1.29}$ if we ignore the $r_\mathrm{h}/r_\mathrm{GW}$ term in Eq.~(\ref{mdeficit}). The dependence on the mass ratio is mild. Results for $t_\mathrm{cusp}$ are shown in the left panel of Figure~\ref{tcusp}. The red curves are for $V_\mathrm{k}=0$, whereas the blue ones assume $V_\mathrm{k}=0.6V_\mathrm{esc}$. If we ignore the $V_\mathrm{k}$ contribution, we can fit the cusp regrowth time as
\begin{equation}
  t_\mathrm{cusp} \approx  6 \left(\frac{M}{10^6 \msun}\right)^{1.19} q^{0.35}~\mathrm{Gyr}.
  \label{tfit}
\end{equation}
The slightly weaker dependence on $M$ than the initial approximation is due to the $r_\mathrm{h}/r_\mathrm{GW}$ term in Eq.~(\ref{mdeficit}). 
Typical cusp regrowth timescales are a significant fraction of the Hubble time for equal-mass binaries with total mass $10^6\msun$, whereas they tend to become unimportant for lower mass MBHs 
(generally less than $1~\mathrm{Gyr}$ for a $10^5\msun$ MBH).

Further core scouring following significant kicks will make these timescales a factor of $2$ longer. For typical kick velocities of few hundred $\mathrm{km\,s}^{-1}$ we find that the EMRI rate drops by a factor of $\sim 2$ due to a combination of MBH ejections from low mass halos and prolongation of cusp regrowth timescales. 

\begin{figure*}
\centering
\includegraphics[width=5.5cm,clip=true,angle=0]{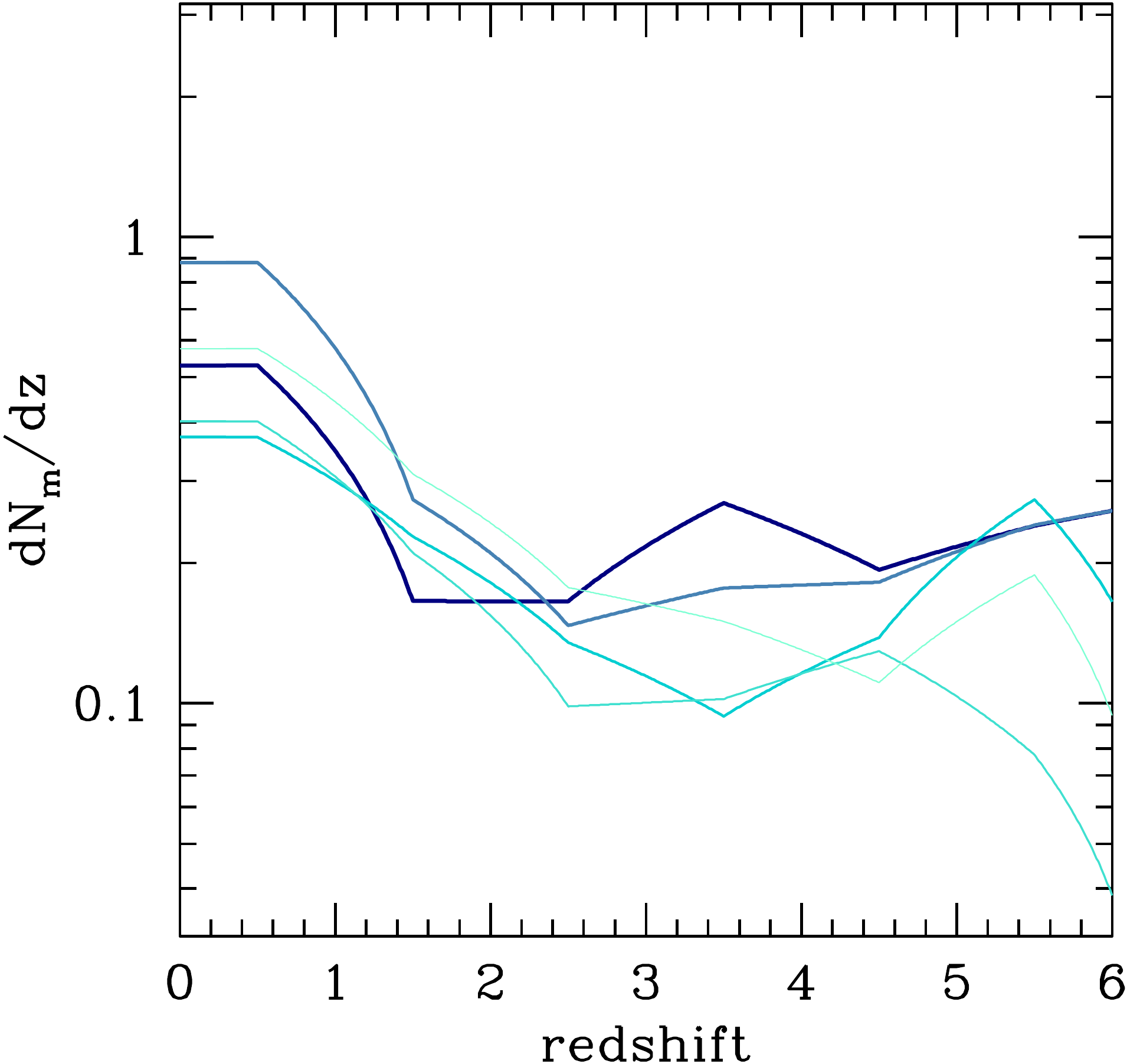} \quad 
\includegraphics[width=5.5cm,clip=true,angle=0]{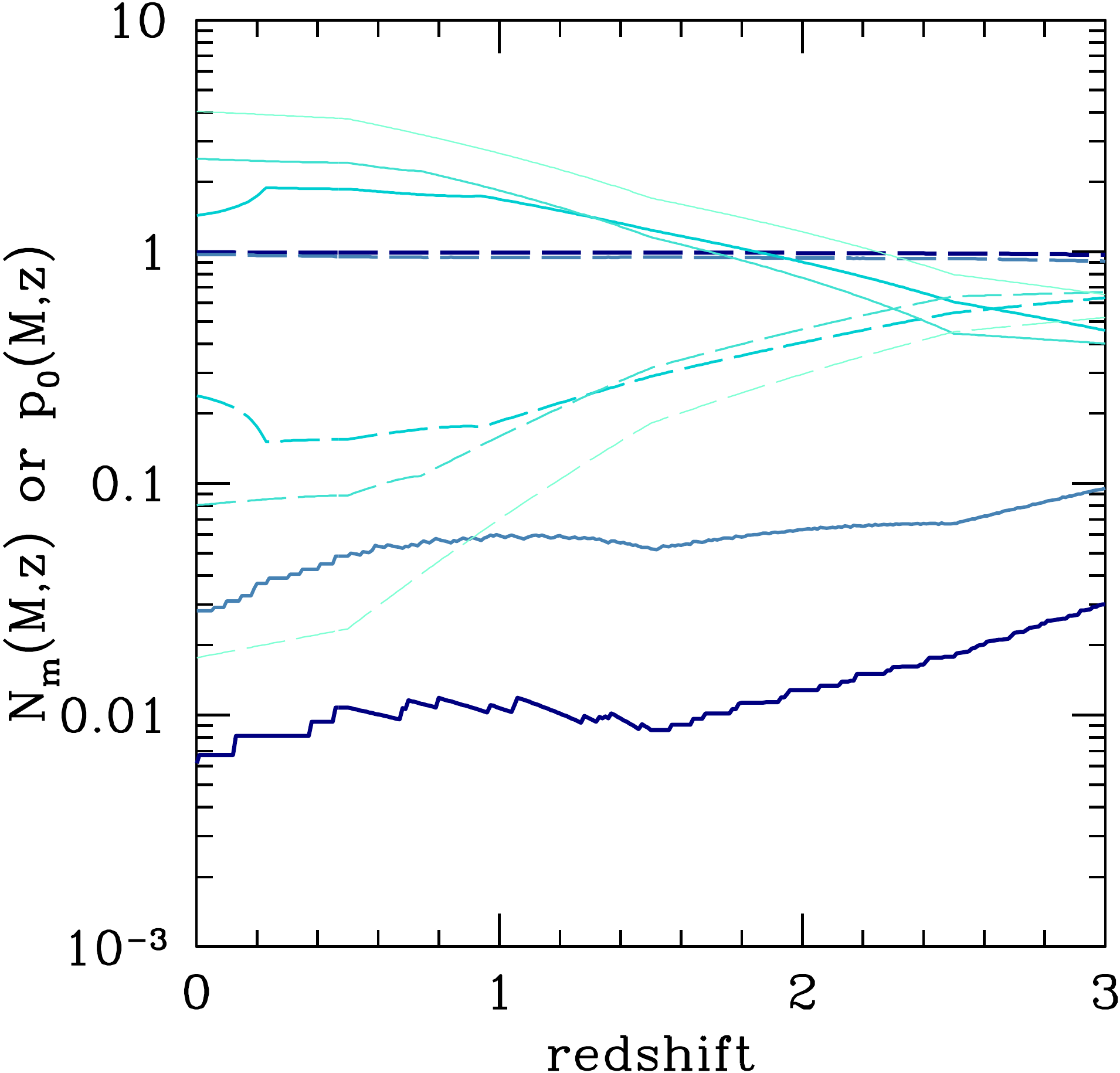} \quad 
\includegraphics[width=5.5cm,clip=true,angle=0]{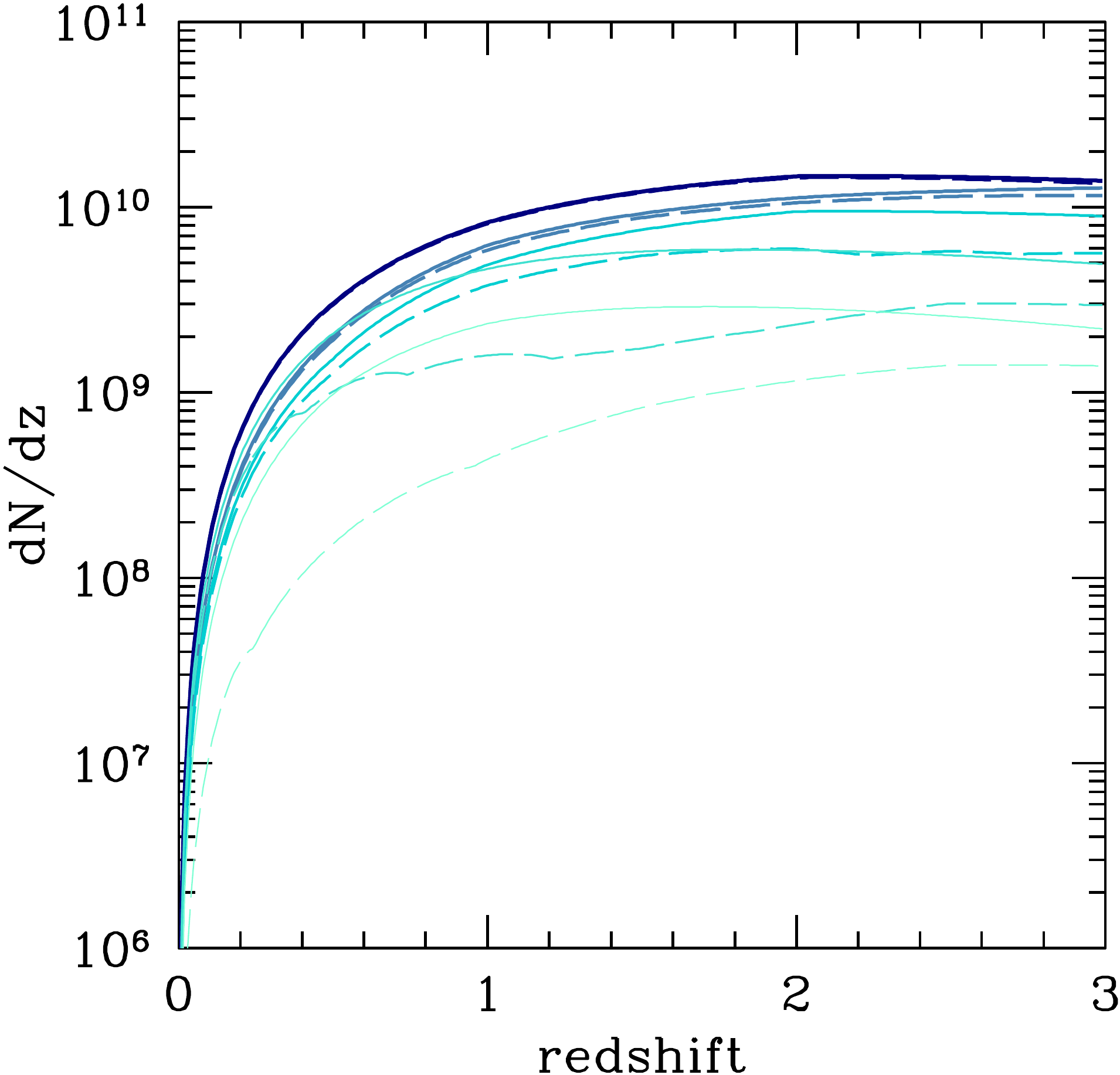}
\caption{Cusp regrowth effect for the popIII model. Left panel: The average differential number of mergers per unit redshift 
(i.e.\ Eq.~(\ref{eq:nmergsingle}) integrated over $q$) $\dd N_m/\dd z$ experienced by each individual MBH of mass $\log_{10} M=4.5,5,5.5,6,6.5$ from darker-thicker to lighter-thinner. Center panel: The solid curves are the values of $N_\mathrm{m}(M,z)$ given by Eq.~(\ref{nmeanmerg}), and the dashed curves are the corresponding
probabilities of retaining a cusp given by Eq.~(\ref{probcusp}). Right panel: The differential number of MBHs $\dd N/\dd z$ across the Universe in the three different mass bins that are potential EMRI hosts, either ignoring cusp disruption (solid lines) or taking it into account (dashed lines).}
\label{distwithcusp2}
\end{figure*}

The adopted $M$--$\sigma$ relation has a significant impact on $t_\mathrm{cusp}$. We will
therefore explore different prescriptions. As mentioned above, our default model employs the $M$--$\sigma$ relation of G{\"u}ltekin et al.~\cite{2009ApJ...698..198G} (labeled ``Gultekin09'') which 
gives $t_\mathrm{cusp}\approx6~\mathrm{Gyr}$ for a Milky Way-like MBH. We also consider two alternatives; a pessimistic model from Kormendy and Ho~\cite{2013ARA&A..51..511K} (labeled ``KormendyHo13'')
which gives $t_\mathrm{cusp}\approx10~\mathrm{Gyr}$ for a Milky Way-like MBH, and an optimistic one~\cite{2013ApJ...764..151G} (labeled ``GrahamScott13'') which gives $t_\mathrm{cusp}\approx2~\mathrm{Gyr}$ for a Milky Way-like MBH.

For the sake of completeness, we also tried a model based on Shankar et al.~\cite{2016MNRAS.460.3119S}, which claims that the observed $M$--$\sigma$ relations are fundamentally biased and that the intrinsic one has a lower normalization. We found this to make little difference in practice for EMRI rates, and do not present results based on this model.

To verify our simple model, we performed a series of sanity checks. First, for a Milky Way-like MBH, Eq.~(\ref{trelax}) implies $t_\mathrm{relax}\approx 10^{11}~\mathrm{yr}$,
which is consistent with Figure~1 of~\cite{2006RPPh...69.2513M}. Moreover, $M_\mathrm{d}$ given by Eq.~(\ref{mdeficit}) is consistent with the results of full $N$-body simulations by 
Khan et al.~\cite{2012ApJ...749..147K}, as shown in the right panel of Figure~\ref{tcusp}. 
Here, each of the red lines shows the mass predicted by our simple model as a function of $q$ (different line styles 
refer to different MBH masses). The blue and green dots are mass deficits computed by Khan et al.~\cite{2012ApJ...749..147K} 
at the end of their simulations. The blue dots are mass deficits within $1.5r_\mathrm{i}$, where $r_\mathrm{i}$ is the MBH binary 
influence radius, whereas the green dots are mass deficits within $3r_\mathrm{i}$. The mass deficit in those simulation saturate between
$2r_\mathrm{i}$ and $3r_\mathrm{i}$. The cores predicted by our simple model are $\approx1.5r_\mathrm{i}$. In the simulations, however, the MBH binaries
do not evolve all the way through coalescence. Small $q$ binaries, in particular, are stopped at an earlier stage of the evolution, because the simulations are more time consuming. This is the likely explanation of the steeper mass-ratio dependence of the simulation results with respect to our models. Overall, the analytical mass deficits and the results of the simulations agree to within a factor of $2$.

\subsubsection{Fraction of MBHs hosted in stellar cusps}

To compute the fraction of MBHs that reside in cusps versus those in cores, we need to convolve the MBH number density $\dd n/\dd M$ (ignoring the
spin dependence for the moment) with the number density of mergers per unit mass, mass ratio and redshift $\dd^3n_\mathrm{m}/\dd M\,\dd z\,\dd q$, and the cusp
regrowth time $t_\mathrm{cusp}(M,q)$ given by Eq.~(\ref{tfit}).

First, we assume that MBHs do not grow appreciably in mass in the redshift range of interest (mostly $z<2$ for LISA). Although this might well be 
a crude approximation, it simplifies the model. From our semianalytic MBH evolution model~\cite{2012MNRAS.423.2533B,2014ApJ...794..104S,2015ApJ...806L...8A,2015ApJ...812...72A}, 
we extract the distribution $\dd^3n_\mathrm{m}/\dd M\,\dd z\,\dd q$,
which is the differential number density (per $\mathrm{Mpc^{3}}$) of mergers with mass ratio $q$ undergone by a MBH of a given mass $M$ at redshift $z$. The 
quantity of interest is $p_0(M,z)$, the probability that a MBH of mass $M$ observed at redshift $z$ had zero mergers within its cusp regrowth time 
$t_\mathrm{cusp}(M,q)$, given by Eq.~(\ref{tfit}). 
We can define the quantity $\dd^2N_\mathrm{m}/\dd z\,\dd q$ as
\begin{equation}
  \frac{\dd^2N_\mathrm{m}}{\dd z\,\dd q}(M,z,q)= \frac{\dd^3n_\mathrm{m}}{\dd M\,\dd z\,\dd q}\left({\diff{n}{M}}\right)^{-1}.
  \label{eq:nmergsingle}
\end{equation}
This is the (mean) differential merger rate for an individual MBH with mass $M$, i.e.\ the number of mergers a MBH of a given mass $M$ has undergone between 
redshift $z$ and $z + \dd z$ and with mass ratio in the range $q$ and $q + \dd q$. The integral over $q$ of Eq.~(\ref{eq:nmergsingle}) is represented in the left panel of Figure ~\ref{distwithcusp2}; 
MBHs in the mass range of interest for LISA generally experience between $0.1$ and $1$ merger per unit redshift since $z=6$. For each mass ratio, we can then define a critical redshift $z_\mathrm{cusp}(M,q)$ by solving the implicit equation 
\begin{equation}
  t_\mathrm{cusp}(M,q)=\int_z^{z_\mathrm{cusp}(M,\,q)}\dd z' \diff{t}{z'},
  \label{tcrit}
\end{equation}
where $t_\mathrm{cusp}(M,q)$ is computed using Eq.~(\ref{trelax}) and Eq.~(\ref{tcuspr}); if a MBH had suffered a merger between $z$ and $z_\mathrm{cusp}(M,q)$, 
there would be no cusp. We can then compute the mean number of mergers $N_\mathrm{m}(M,z)$ experienced by an individual MBH of mass $M$ observed at redshift $z$ in its cusp regrowth time as 
\begin{equation}
  N_\mathrm{m}(M,z)=\int \dd q \int_z^{z_\mathrm{cusp}(M,\,q)}\dd z' \frac{\dd^2 N_\mathrm{m}}{\dd z'\,\dd q}(M,z,q).
  \label{nmeanmerg}
\end{equation}
Assuming Poissonian statistics for the mergers, the probability that a MBH of mass $M$ and redshift $z$ did not suffer a merger within its cusp 
regrowth time is\footnote{A Poissonian probability distribution is strictly speaking valid only for rare, statistically independent events with a constant rate per unit time. Nevertheless, one can easily show that
this equation holds also for events with non-constant rate. 
}
\begin{equation}
  p_0(M,z)= \exp\left[-N_\mathrm{m}(M,z)\right].
  \label{probcusp}
\end{equation}
We apply to each MBH a probability $p_0(M,z)$ of being hosted in a stellar cusp, therefore being a suitable candidate for capturing an EMRI. $N_\mathrm{m}(M,z)$ 
is shown in the central panel of Figure~\ref{distwithcusp2} for different MBH mass values. Despite the similar number of mergers across the mass spectrum 
(left panel), low mass MBHs observed at any $z$ are extremely unlikely to have undergone a merger within their short cusp regrowth timescale (see left panel of 
Figure~\ref{tcusp}), and their probability of being hosted in a stellar cusp is of order unity. The opposite is true for massive MBHs which reside in galaxies with 
much longer cusp regrowth timescales, and are likely to be hosted in a low-density stellar core.

If the distribution of MBHs in the Universe is described by a mass function (now including spin) of the form $\dd^3N/(\dd M\,\dd z\,\dd a)$, 
then we can construct an effective MBH mass function for MBHs which could be potential EMRI hosts:
\begin{equation}
  \left(\frac{\dd^3N}{\dd M\,\dd z\,\dd a}\right)_\mathrm{eff}=\frac{\dd^3N}{\dd M\,\dd z\,\dd a} \, p_0(M,z).
  \label{effectivemf}
\end{equation}
The right panel of Figure~\ref{distwithcusp2} shows this distribution integrated in spin and in different mass bins. 
It is clear that the number of potential EMRI hosts is severely suppressed only for $M>10^6\msun$.

\subsection{EMRI rate per MBH and properties of the stellar-mass BH}

Finally, we need to specify the rate $R_0$ at which COs are captured by the central MBH, and define the properties of their orbits. 
The CO capture rate by MBHs has been investigated extensively in the literature, taking into account the effect of mass 
segregation~\cite{2011CQGra..28i4017A}, resonant relaxation~\cite{2006ApJ...645.1152H}, relativistic corrections~\cite{2011PhRvD..84d4024M}, central 
MBH spins~\cite{2013MNRAS.429.3155A} and initial density profiles of the COs~\cite{2015ApJ...814...57M}. 

Our starting point is the intrinsic rate from Amaro-Seoane and Preto~\cite{2011CQGra..28i4017A}, which accounts for the effect of mass segregation:
\begin{equation}
  R_0 = 300 \left( \frac{M}{10^6\msun} \right)^{-0.19}~\mathrm{Gyr}^{-1}.
  \label{ratesingle}
\end{equation}
This is is useful scaling relation; however, it has been calibrated for Milky Way-like galaxies, and care must be taken when extrapolating 
to other systems. In particular, this rate was calculated assuming a steady-state stellar environment 
surrounding the (growing) MBH which often cannot be achieved, especially for low-mass MBHs.
Moreover, Eq.~(\ref{ratesingle}) only describes the EMRI rate: it does not include \emph{direct plunges}. 
COs can be scattered onto nearly radial orbits, directly plunging into the MBH without emitting a 
significant GW signal. Although such systems are lost as GW sources, they do contribute to the growth of the 
MBH. The ratio of plunges to EMRI depends mostly on the steepness of the density profile of the CO population. 
Compared to EMRIs, plunges are typically scattered into the MBH from much greater distances, so that a flatter 
density profile results in a larger plunge-to-EMRI ratio. For example, Merritt~\cite{2015ApJ...814...57M} considered two 
different CO distributions around MBHs of $10^6\msun$ and $4\times10^6\msun$, and found that while the EMRI rate varied within a 
factor of $2$, remaining consistent with Eq.~(\ref{ratesingle}), the number of plunges per EMRI, $N_p$, went from being less than one for 
the steeper density profile, to be more than $50$ for the shallower one. A recent study including a single population of compact objects 
found more than $100$ plunges per EMRI~\cite{2016ApJ...820..129B}.
A proper computation of EMRI rates in an astrophysical context would require $N$-body simulations starting from realistic initial 
conditions, spanning a wide range of MBH masses and of their surrounding stellar distribution properties. This is a challenge that 
goes beyond the scope of this paper, and in the following we develop a simple model to quantify the impact of non-stationary CO feeding 
rates and direct plunges on the astrophysical EMRI rates.

The parameter $N_\mathrm{p}$ introduced earlier can vary between zero and $\sim10^2$. 
Using Eq.~(\ref{ratesingle}), the total mass accretion rate for the MBH is given by
\begin{align}
  \dot{M} &= m R_0(1+N_\mathrm{p}) \nonumber\\
          &= 3000(1+N_\mathrm{p})\left( \frac{m}{10\msun} \right)\left( \frac{M}{10^6\msun} \right)^{-0.19}~\msun\,\mathrm{Gyr}^{-1},
  \label{maccrate}
\end{align}
where $m$ is the characteristic mass of the CO. There are two problems that arise from implementing Eq.~(\ref{maccrate}), which 
are exacerbated for low MBH masses. Consider for example $M=10^5\msun$, $m=10\msun$ and $N_\mathrm{p}=10$. First, according to 
this prescription, the MBH would double its mass in only $2~\mathrm{Gyr}$, accreting more than five times its initial mass in a
Hubble time. Therefore, accreting COs at the rate given by Eq.~(\ref{ratesingle}) would be inconsistent with the existence of $M=10^5\msun$ MBHs~\cite{2017arXiv170100415A}. 
Second, such a high accretion rate implies an astrophysically implausible supply of COs to the MBH. Assuming a standard Salpeter mass function~\cite{1955ApJ...121..161S,2001MNRAS.322..231K}, only about $0.3\%$ of stars have a mass $m_*>30\msun$. 
Assuming those end their life as COs of $m\approx10\msun$~\cite{2002RvMP...74.1015W,2015MNRAS.451.4086S}, 
then we can estimate that about $3\%$ of the total stellar bulge is indeed composed by COs. Within the sphere of influence of 
the MBH the enclosed mass in stars is $M_*=2M$, and therefore the mass in remnant BHs is about $M_\mathrm{CO}=0.06M$. 
The CO content within the sphere of influence of the MBH would therefore be depleted in a time
\begin{align}
  t_\mathrm{d} &=\frac{M_\mathrm{CO}}{\dot{M}}=\frac{0.06M}{m\,R_0(1+N_\mathrm{p})} \nonumber\\
      &=\frac{20}{1+N_\mathrm{p}}\mathrm{Gyr}\left( \frac{m}{10\msun} \right)^{-1}\left( \frac{M}{10^6\msun} \right)^{1.19}.
  \label{tdepletion}
\end{align}
This can be compared to the relaxation time defined by Eq.~(\ref{trelax}), where we can substitute $r_\mathrm{c}$ 
with the influence radius $r_\mathrm{i}\approx2GM/\sigma^2$ of the central MBH. By using the $M$--$\sigma$ relation 
of G{\"u}ltekin et al.~\cite{2009ApJ...698..198G} and assuming $\ln\Lambda=10$, we can compute the ratio of the two timescales:
\begin{equation}
\frac{t_\mathrm{d}}{t_\mathrm{relax}} \simeq \frac{1.2}{1+N_\mathrm{p}}\left( \frac{m}{10\msun} \right)^{-1}\left( \frac{M}{10^6\msun} \right)^{0.06}.
  \label{tratio}
\end{equation}
Although Eq.~(\ref{tratio}) is valid for an isothermal density profile and employed\ a specific $M$--$\sigma$ 
relation~\cite{2002ApJ...574..740T}, we verified that for more sophisticated Dehnen profiles~\cite{1993MNRAS.265..250D}
and alternative scaling relations the result holds within a factor of $2$. The ratio is roughly 
independent of mass (but does depend on the adopted $M$--$\sigma$ relation), and most importantly, it
is larger than unity only if $N_\mathrm{p}\approx 0$. In this case, the depletion time is longer than the relaxation time and we can therefore assume that the EMRI rate is sustainable.
However, since in general there are several plunges per EMRI, ${t_\mathrm{d}}/{t_\mathrm{relax}}<1$ and a steady state situation
where the EMRI rate is given by Eq.~(\ref{ratesingle}) cannot be sustained. We therefore define a duty cycle
\begin{equation}
  \Gamma = \min \left\{ \frac{t_\mathrm{d}}{t_\mathrm{relax}} , 1\right\} ,
  \label{eq:gammacorr}
\end{equation}
and a sustainable EMRI rate is given by $\Gamma R_0$.

\begin{figure}
\centering
\includegraphics[width=\columnwidth,clip=true,angle=0]{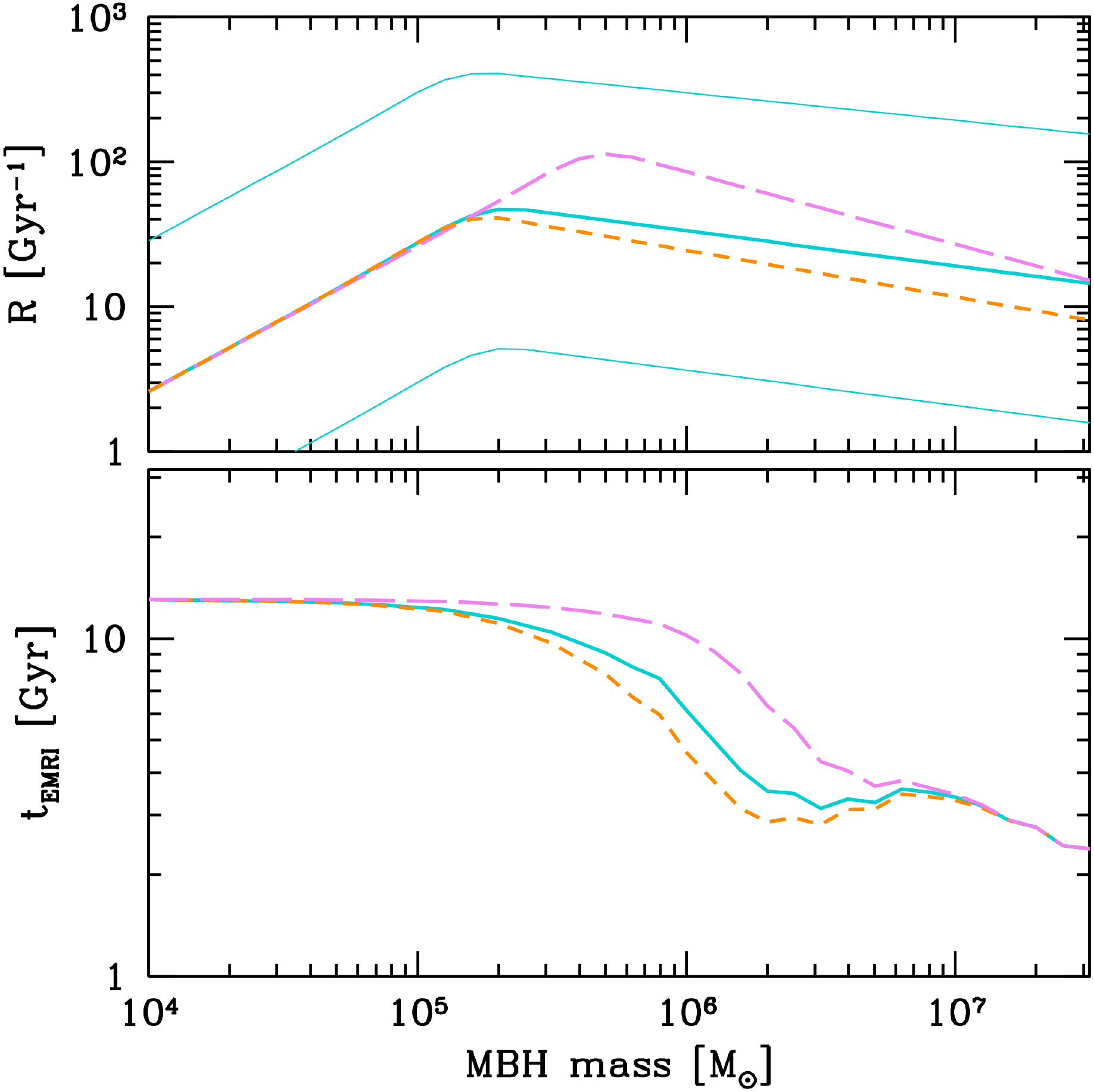}
\caption{Top panel: The adjusted EMRI rate computed 
according to Eq.~(\ref{correctedrate}). The three (central) thick lines assume $N_\mathrm{p}=10$ and
correspond to the pessimistic (KormendyHo13, short-dashed orange), 
fiducial (Gultekin09, solid turquoise) and optimistic (GrahamScott13, long-dashed violet) $M$--$\sigma$ 
relations. The two thin turquoise lines show the rates for the
fiducial model, but assuming $N_\mathrm{p}=0$ (lower curve) and $N_\mathrm{p}=100$ (upper curve).
Lower panel: The average time $t_\mathrm{EMRI}$ that a MBH of a given mass is surrounded by a stellar cusp, 
and is therefore a potential EMRI source, as implicitly defined by Eq.~(\ref{temri}). 
The curves are for the same three different $M$--$\sigma$ 
relations in the top panel.}
\label{fig:temri}
\end{figure}

We can now compute a MBH mass growth by combining this rate with the amount of time a given MBH is surrounded by a cusp, and is therefore a potential EMRI host. We define this time as
\begin{equation}
  t_\mathrm{EMRI}=\int \dd z \,\diff{t}{z} p_0(M,z),
  \label{temri}
\end{equation}
where $p_0(M,z)$ is given by Eq.~(\ref{probcusp}) and represents the probability that a MBH of a particular mass is hosted in a stellar cusp as a function of redshift. This time is plotted in the lower panel of Figure~\ref{fig:temri} and, as expected, is essentially the Hubble time $T_H$ at $M<10^5\msun$ and rapidly drops to $2~\mathrm{Gyr}$ at $M>10^6\msun$.
The mass growth is then
\begin{equation}
  \Delta{M}= m \Gamma {R}_0t_\mathrm{EMRI}.   
  \label{deltam}
\end{equation}

Even with the corrective factor of Eq.~(\ref{eq:gammacorr}), the supply of COs on inspiralling and plunging orbits can overgrow MBHs. If, for example, $t_\mathrm{EMRI}=T_H$ and $m=10\msun$ is the mass of the accreted CO, then from Eq.~(\ref{deltam}) $\Delta{M}= m \Gamma {R}_0 T_H \gtrsim M$ for $M\approx 10^4\msun$. For a practical computation of the rate we therefore introduce a damping factor to (arbitrarily) cap the maximum allowed mass growth to be $\exp(-1) M$, so that the MBH can at most grow by an $e$-fold due to CO accretion in its lifetime. We pick $N_\mathrm{p}$ and compute $\Gamma$ from Eq.~(\ref{tratio}) considering a specific $M$--$\sigma$ relation, and then calculate, for each MBH mass, $\Delta{M}$ from Eq.~(\ref{deltam}). Using this, the damping factor is defined as 
\begin{equation}\label{eq:kappa}
\kappa = \min \left\{ \exp(-1)\frac{M}{\Delta{M}} , 1 \right\}.
\end{equation}
%
%
Incorporating this, the effective EMRI rate is given by:
\begin{equation}
  R= \kappa\,\Gamma R_0.
\label{correctedrate}
\end{equation}

Examples of the impact of the factors $\Gamma$ and $\kappa$ on the EMRI rate $R$ are shown in the upper panel of Figure~\ref{fig:temri}. 
From Eq.~(\ref{tdepletion}), it is clear that the value of the $\Gamma$ factor, and hence $R$, depends critically on $N_\mathrm{p}$. 
We therefore explore three different models featuring $N_\mathrm{p}=0$, $10$, and $100$. Since $\Gamma\approx 1$ for $N_\mathrm{p}=0$, $N_\mathrm{p}$ 
has the obvious effect of renormalizing the EMRI rate. Eq.~(\ref{tratio}) implies that $R_0$ given by Eq.~(\ref{ratesingle}) is close to the supply
CO rate allowed by relaxation; if a large fraction of those COs result in direct plunges rather than EMRIs, the EMRI rate must drop accordingly.
Different $M$--$\sigma$ relations alter the slope of the rate as a function of mass because of its influence on the relaxation time and so $\Gamma$. 
The $\kappa$ factor affects the rates mostly for masses below $\sim10^5 \msun$, where overgrowth by CO accretion is easy. This makes our EMRI estimates conservative since it implies a rate suppression. We will see later that most of LISA detections come from systems with $M>10^{5}\msun$, and therefore our results are not severely impacted by the introduction of this damping factor.

\subsubsection{Eccentricity and inclination}
\label{sec:ecc}
To estimate the distribution of EMRI eccentricities at the last stable orbit (LSO), we evolved a large sample of COs from their typical capture radius (which is of the order of $0.01~\mathrm{pc}$ for a $10^6\msun$ central MBH~\cite{2007CQGra..24R.113A}) to final plunge. We found a rather flat eccentricity distribution at plunge in the range $0<e_\mathrm{p}<0.2$, with a small tail of outliers with higher $e_\mathrm{p}$. In the following, we therefore assume a flat distribution in the range  $0<e_\mathrm{p}<0.2$ for simplicity.

Finally, the rates in Eq.~(\ref{ratesingle}) have been computed in the nonspinning approximation. As a consequence of frame-dragging effects, the location of the LSO of a test particle orbiting a spinning MBH depends on the inclination of its orbital angular momentum with respect to the MBH spin, $\theta_\mathrm{inc}$, and on whether the orbit is prograde ($0 \leq \theta_\mathrm{inc} \leq \pi/2$) or retrograde ($\pi/2 \leq \theta_\mathrm{inc} \leq \pi$). Therefore, the rate $R$ has to be adjusted using a spin-dependent and inclination correcting factor $W(a,\theta_\mathrm{inc})$~\cite{2013MNRAS.429.3155A}.  The correction factor is based on the ratio of the semimajor axis for the spinning case $a^\mathrm{Kerr}_\mathrm{LSO}$ with respect to the nonspinning case $a^\mathrm{Schw}_\mathrm{LSO}$ averaged over the eccentricity, that is~\cite{2013MNRAS.429.3155A}:  
\begin{equation}
W(a,\theta_\mathrm{inc}) = \left\langle \frac{a^\mathrm{Kerr}_\mathrm{LSO}(e)}{a^\mathrm{Schw}_\mathrm{LSO}(e)} \right\rangle_{e} \,.
\label{W-factor}
\end{equation}
In practice we use an averaged correcting factor $W(a)$ which is the result of averaging over the orbital inclination
\begin{equation}
W(a) = \left\langle W(a,\theta_\mathrm{inc}) \right\rangle_{\theta_\mathrm{inc}} \,.
\label{W-factor-averaged}
\end{equation}
In terms of this function, the event rates for EMRIs in the spinning case are related to the nonspinning approximation by
\begin{equation}
R^\mathrm{Kerr}(a) = R^\mathrm{Schw}\; [W(a)]^{-0.83}\,,
\end{equation}
assuming an old, segregated cusp of COs of mass $10\,M_{\odot}$ around the MBH~\cite{2009ApJ...697.1861A,2011CQGra..28i4017A}.

\subsection{Putting the pieces together}
\label{sec:bakery}

In summary, the EMRI rate depends on a number of ingredients, as we described above:
\begin{enumerate}
\item The MBH mass function, for which we assume two models: Barausse12 and Gair10.
\item The MBH spin distribution, for which we explore three cases: the near-maximally spinning distribution (a98); a flat spin distribution (aflat), and nonspinning MBHs (a0).
\item The $M$--$\sigma$ relation, defining the properties of the stellar distribution surrounding the MBH,
the cusp regrowth time following MBH binary erosion, and the EMRI duty cycle. We consider 
three relations: Gultekin09, KormendyHo13 and GrahamScott13. We also consider an extra model assuming the Gultekin09 relation but with no cusp erosion.
\item The ratio of plunges to EMRIs, assumed to be $N_\mathrm{p}=0$, $10$ and $100$.
\item The characteristic CO mass, for which we consider both $m=10\msun$ and $m=30\msun$.  
\end{enumerate}  

Our default model is based on the self-consistent semianalytic code for MBH formation and cosmic evolution developed in Barausse~\cite{2012MNRAS.423.2533B}. The MBH mass function 
is therefore the Barausse12, and MBHs are consistently maximally spinning (a98). We use the $M$--$\sigma$ relation Gultekin09 to compute the 
cusp regrowth time following MBHB mergers and the EMRI duty cycle. We assume a moderately large number of plunges per EMRI $N_\mathrm{p}=10$ 
and a characteristic CO mass $m=10\msun$. Starting from this default set-up,
we explore the effect of each single ingredient listed above by varying them individually keeping all the other fixed. We further explore the
most optimistic and pessimistic models allowed by all the combinations of the ingredients listed above.

In total, we consider $12$ models that we label ``M$x$'' with $x=1,\ldots,12$. The default setup described above is indicated as M1, 
and the key to read the models and their main properties are listed in Table~\ref{tab1}.

\begin{table*}
\begin{ruledtabular}
\centering
\begin{tabular}{ccccccc|ccc}
\squeezetable
 & Mass &  MBH & Cusp & $M$--$\sigma$ & & CO & & EMRI rate [$\mathrm{yr}^{-1}$] & \vspace{-0.04in}\\
Model &  function &  spin &  erosion &  relation & $N_\mathrm{p}$ &  mass [$\msun$] & Total  & Detected (AKK)  & Detected (AKS) \\
\hline
M1 & Barausse12 & a98   & yes & Gultekin09    & 10  & 10 & 1600 &  294& 189\\ 
M2 & Barausse12 & a98   & yes & KormendyHo13  & 10  & 10 & 1400 &  220& 146\\ 
M3 & Barausse12 & a98   & yes & GrahamScott13 & 10  & 10 & 2770 &  809& 440\\ 
M4 & Barausse12 & a98   & yes & Gultekin09    & 10  & 30 &  520 (620) & 260 &221\\ 
M5 & Gair10     & a98   & no  & Gultekin09    & 10  & 10 &  140 &  47& 15\\ 
M6 & Barausse12 & a98   & no  & Gultekin09    & 10  & 10 & 2080 &  479& 261\\ 
M7 & Barausse12 & a98   & yes & Gultekin09    & 0   & 10 & 15800 &  2712& 1765\\ 
M8 & Barausse12 & a98   & yes & Gultekin09    & 100 & 10 &  180 &  35& 24\\ 
M9 & Barausse12 & aflat & yes & Gultekin09    & 10  & 10 & 1530 &  217& 177\\ 
M10 & Barausse12 & a0    & yes & Gultekin09    & 10  & 10 & 1520 &  188& 188\\ 
M11 & Gair10     & a0    & no  & Gultekin09    & 100 & 10 &   13 &  1& 1\\ 
M12 & Barausse12 & a98   & no  & Gultekin09    & 0   & 10 & 20000 &  4219& 2279\\ 
\end{tabular}
\caption{
List of EMRI models considered in this work. Column 1 defines the label of each model. For each model we specify the MBH mass function (column 2), the MBH spin model (column 3), whether we consider the effect of cusp erosion following MBH binary mergers (column 4), the $M$--$\sigma$ relation (column 5), the ratio of plunges to EMRIs (column 6), the mass of the COs (column 7); the total number of EMRIs occurring in a year up to $z=4.5$ (column 8; for model M4 we also show the total rate per year up to $z=6.5$); the detected EMRI rate per year, with AKK (column 9) and AKS (column 10) waveforms. The AKK and AKS waveforms are introduced in Section~\ref{sec:analysis}, and bracket waveform modelling uncertainties.}
\label{tab1}
\end{ruledtabular}
\end{table*}

For each model we construct the population of EMRIs by Monte Carlo sampling from the distribution $\dd^3N/(\dd M\,\dd z\,\dd a)\times p_0(M,z) R(M,a)$. 
This gives a catalog of EMRIs including the two masses $(M,m)$, redshift of the event $z$, and MBH spin $a$. 
To define each individual event and construct EMRI waveforms we need to specify $10$ more parameters: 
\begin{itemize}
\item Phase, sky position and orientation angles: we assume that the sky position and spin orientation vectors are distributed isotropically on the sphere. The three phases at plunge corresponding to orbital phase, phase of precession of the periapsis and phase of precession of the orbital plane are uniformly distributed between $0$ and $2\pi$.
\item Inclination and eccentricity are distributed as described in Section~\ref{sec:ecc}.
\item Plunge times are taken to be uniform in $[0,2]~\mathrm{yr}$. We ignore events that plunge after the end of the mission duration, although they might be detectable if they are close enough.
\end{itemize}

Table~\ref{tab1} illustrates the potential range in the intrinsic EMRI rate.
The last column lists the 
number of EMRIs occurring in the Universe in $1$~year (observed at Earth) up to $z=4.5$ (for model M4 we also report the rate up to $z=6.5$ in parentheses).

Numbers span more than three order of magnitudes, ranging from about $10$ to $20000$. The variation is mostly due to the unknown number of plunges and to the poorly constrained MBH mass function at $M<10^6\msun$. Cusp erosion has a relative minor effect on the rates (a factor of $2$).\footnote{This could be up to a factor of $4$ if kick velocities of few hundred $\mathrm{km\,s}^{-1}$ are considered in the computation of the cusp regrowth timescale (cf.\ Eq.~\ref{mdeficit}).} Even smaller is the effect of spin, affecting EMRI rates at the $10\%$ level; there are more EMRIs when spins are higher as the LSO is smaller (and so it is more difficult to directly plunge~\cite{2013MNRAS.429.3155A}), but this only affects a small portion of orbits. However, we will see that spins will play a more important role in the detectability of these events by LISA. Changing the $M$--$\sigma$ relation, which sets the relation between the MBH and its surrounding population of COs, can introduce a variation of about a factor of $2$. More significant are the mass of the COs and the number of plunges, as both of these directly impact the mass accreted by the MBH and so the necessary duty factor to preserve the population of MBHs. An increase in either $m$ or $N_\mathrm{p}$ by a factor of $X$ reduces the EMRI rate by a similar factor. Since we are more uncertain of the number of plunges, this  has  a greater potential impact on the expected rate, here changing it by almost two orders of magnitude. A drop of about one order of magnitude is achieved by switching to the pessimistic MBH mass distribution, as the reduction in the number of MBHs naturally decreases the number of EMRIs.


For each of the $12$ models outlined above we generate $10$ Monte-Carlo realizations of the expected population of EMRIs plunging in $1$~year.
We therefore construct a library of $120$ catalogs that includes all EMRI events occurring in the Universe in $10$~years for the $12$ models. 

\section{Waveforms, signal analysis and parameter estimation}
\label{sec:analysis}
Having generated astrophysical populations of EMRI systems, we need to determine which of the systems will be observed by LISA. To do this, we need a model of the GW emission from an EMRI system. 
Accurate gravitational waveforms from EMRIs can be computed using BH perturbation theory, exploiting 
the large difference in masses of the two objects to regard the smaller as a perturbation of the spacetime of the larger and construct an expansion in the mass ratio (see~\cite{2011LRR....14....7P} for a review). Perturbative calculations have not yet been completed to the order necessary to accurately track the phase of an EMRI over an entire inspiral, and these calculations are extremely computationally expensive. Two approximate EMRI models have therefore been developed, which capture the main features of EMRI waveforms at much lower computational cost and can therefore be used to explore questions connected to the detection and scientific exploitation of EMRI observations. Of the two models, the numerical kludge~\cite{2006PhRvD..73f4037G,2007PhRvD..75b4005B} is the more accurate 
and is based on modelling the trajectory of the smaller object as a geodesic of the Kerr background, with inspiral imposed on the system. With further enhancements, 
the numerical-kludge model may be accurate enough for use in LISA data analysis. However, it is still relatively computationally expensive. The analytic kludge (AK) model~\cite{2004PhRvD..69h2005B} 
is computationally cheaper, at the cost of less faithfulness to real EMRI signals. The AK model approximates gravitational wave emission by that from a Keplerian orbit~\cite{1963PhRv..131..435P}, with 
precession of the orbital perihelion, precession of the orbital plane, and inspiral of the orbit added using post-Newtonian prescriptions. The AK model provides only an approximation to the true strong-field dynamics, as the orbital frequencies do not exactly match~\cite{2015CQGra..32w2002C}, and features such as the final plunge or transient resonances~\cite{2012PhRvL.109g1102F,2016PhRvD..94l4042B} 
are neglected and cannot be readily incorporated. However, the model is cheap to generate and it should include the most important qualitative features of real EMRI signals. 
The simplicity of the model allows it to be generated in the large numbers required to examine EMRI science questions such as those being explored in this paper, and so we use 
it here. The AK model has been widely used for similar applications in the literature, in particular it was the EMRI model used in the context of the Mock LISA Data Challenges 
(MLDCs)~\cite{2006AIPC..873..625.,2007CQGra..24S.551A,2008CQGra..25r4026B,2010CQGra..27h4009B}.

The AK model is known to be imperfect, and so in order to quantify inaccuracies we consider two different variants. In the classic work by Barack and Cutler~\cite{2004PhRvD..69h2005B}, the AK model was cut off when the orbital frequency reached the value corresponding to the Schwarzschild LSO. We denote this form of the AK model by ``AKS'', where the ``S'' stands for ``Schwarzschild''. Prograde inspirals into spinning MBH can get much closer before plunge, generating many cycles of higher frequency and amplitude. Thus, omitting those cycles from the model is likely to significantly underestimate the possible signal-to-noise ratio (SNR). An alternative is to continue the inspiral until the frequency reaches the Kerr ISCO. We denote this form of the AK model by ``AKK'', where the ``K'' stands for ``Kerr''. The post-Newtonian evolution equations used to construct the AK model are increasingly inaccurate as the orbital separation decreases, and so the additional portion of inspiral included in the AKK model is unlikely to be accurately represented, and most likely will lead to an over-estimate of the SNR. We will present results for both the AKK model and the AKS model in order to quantify the uncertainty that comes from the modelling assumptions. SNRs can also be computed using results from BH perturbation theory, in particular solutions to the Teukolsky equation, which provides the first-order radiative part of the perturbative evolution. Teukolsky results for circular, equatorial inspirals into spinning BHs were presented in Finn and Thorne~\cite{2000PhRvD..62l4021F}, and we can use those results to assess the accuracy of the AKS and AKK prescriptions. 

\begin{figure}
\centering
\includegraphics[width=1.03\columnwidth,clip=true,angle=0]{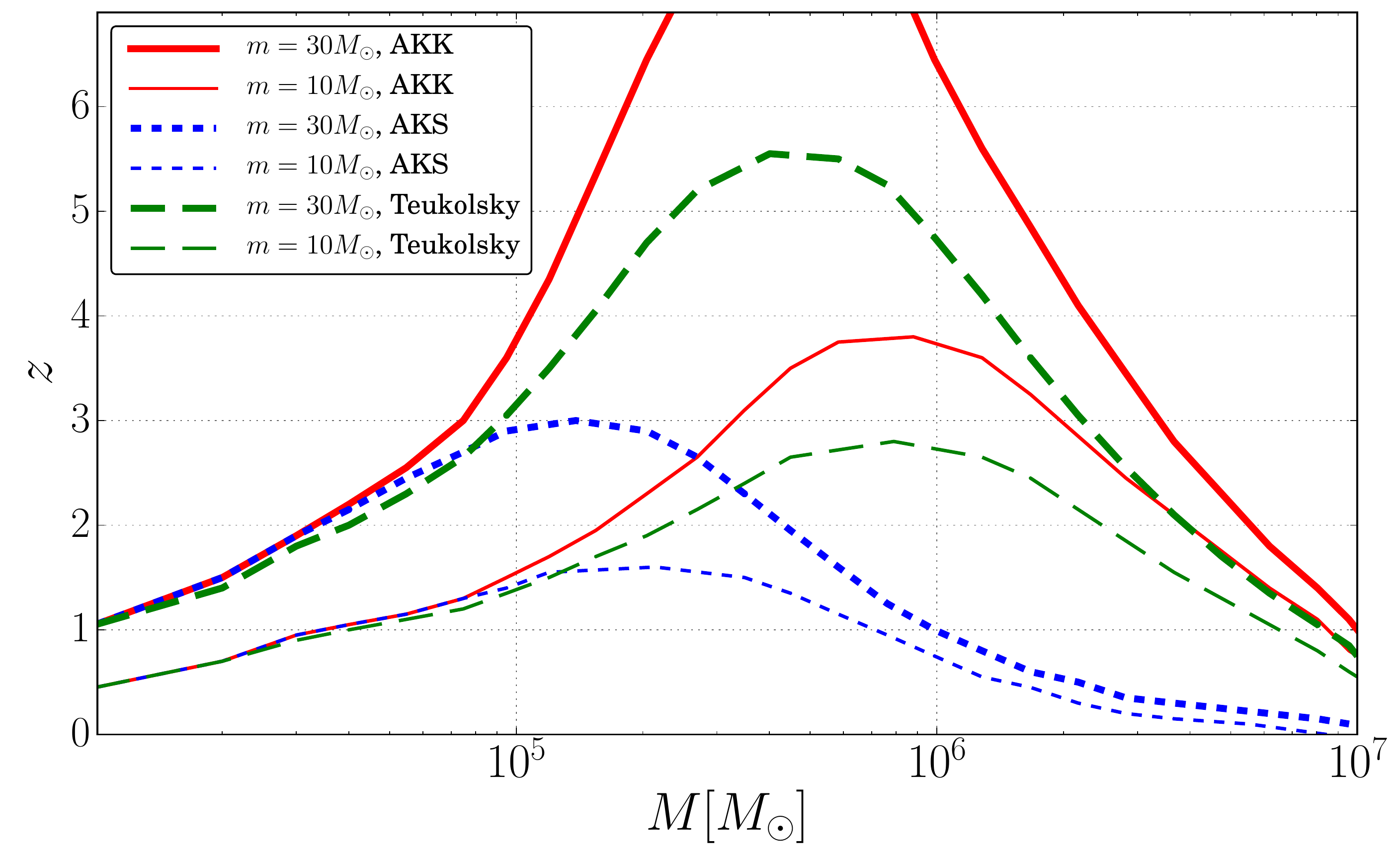}
\caption{The redshift at which the sky-averaged SNR of a prograde, circular, equatorial EMRI into a MBH with spin $a=0.99$ reaches the threshold $\varrho = 20$. The horizon is shown as a function of intrinsic MBH mass and for the two different choices of the compact object mass used in these studies, $m=10M_\odot$ and $m=30M_\odot$. The horizon is computed using accurate Teukolsky fluxes and using a Newtonian inspiral truncated either at the Schwarzschild ISCO, labelled ``AKS'', or at the Kerr ISCO, labelled ``AKK''. Individual sources may be detected to even larger distances if their orientation is near optimal.}
\label{teuk} 
\end{figure}
 
Finn and Thorne~\cite{2000PhRvD..62l4021F} tabulate their results in terms of corrections relative to a Newtonian inspiral. By setting those corrections equal to $1$ we can obtain SNRs for Newtonian inspirals, which we can terminate at the Schwarzschild ISCO or at the Kerr ISCO. This provides an approximation to the AK model, which is built on Newtonian inspirals, albeit with precession added and inspiral augmented by higher order corrections. Figure~\ref{teuk} shows the sky-averaged horizon distance for a prograde, circular, equatorial inspiral into a black hole with spin $a=0.99$, computed either using the Teukolsky fluxes, or using Newtonian inspirals truncated at the two different ISCOs. We see that, as expected, the approximate AKS and AKK horizons bracket the accurate Teukolsky horizon. The AKS horizon suggests increased sensitivity to lower mass black holes, while the AKK horizon has peak sensitivity at the same MBH mass as the Teukolsky horizon. Although these are just approximations to the true AKS and AKK horizons, we expect the true horizons to have the same shape with the AKS horizon extending to slightly higher redshift than the Newtonian calculations indicate and the AKK horizon to slightly lower redshift, still bracketing the true horizon.

Given a waveform model, we represent the sensitivity of LISA to a given EMRI by a simple SNR threshold. If the EMRI has SNR above the specified 
threshold, the system will be detected, otherwise it will not. Early work on EMRIs assumed that an SNR of $30$ would be required for detection, 
to allow for the complexities of LISA data analysis~\cite{2004CQGra..21S1595G}. 
However, in the MLDCs EMRI signals with SNRs as low as $\sim15$ were successfully identified, albeit under idealized 
conditions~\cite{2010CQGra..27h4009B}. Therefore, we use a more modest SNR threshold of $20$. The SNR is calculated as
\begin{equation}
\varrho = \left\langle h \middle| h \right\rangle^{1/2}
\end{equation}
using the noise-weighted inner product~\cite{1992PhRvD..46.5236F}
\begin{equation}
\left\langle g \middle| h \right\rangle = 2 \int_{0}^{\infty}\frac{\tilde{g}(f)\tilde{h}^*(f)+\tilde{g}^*(f)\tilde{h}(f)}{S_n(f)} \, \dd f,
\end{equation}
where the EMRI signal is denoted by $h(t; \boldsymbol{\Theta})$, $\boldsymbol{\Theta}$ represents the parameters of the signal, a tilde indicates the Fourier transform of the signal, and $S_n(f)$ is the noise power spectral density. 

In the limit of suitably high SNR~\cite{2008PhRvD..77d2001V}, the likelihood for the parameters can be approximated as a Gaussian~\cite{1994PhRvD..49.2658C}
\begin{equation}
\mathcal{L}(\boldsymbol{\Theta}) \propto \exp\left(-\frac{1}{2} \sum_{i,\,j}\left\langle \diff{h}{\Theta^i} \middle| \diff{h}{\Theta^j} \right\rangle \Delta\Theta^i \Delta\Theta^j \right),
\end{equation}
where $\Delta \Theta^i$ represents the displacement from the peak of the distribution (which coincides with the true value in this approximation) for the $i$-th parameter. The Fisher matrix has elements
\begin{equation}
\Gamma_{ij} = \left\langle \diff{h}{\Theta^i} \middle| \diff{h}{\Theta^j} \right\rangle,
\end{equation}
and the covariance matrix (the inverse of the Fisher matrix) gives the Cram\'{e}r--Rao bound on the true width of the distribution~\cite{2008PhRvD..77d2001V}. The variance (uncertainty squared) for the $i$-th parameter can be approximated by $\sigma_i^2 = (\Gamma^{-1})_{ii}$.

\section{Results}
\label{sec:results}

\begin{figure*}
\centering
\includegraphics[width=\textwidth,clip=true,angle=0]{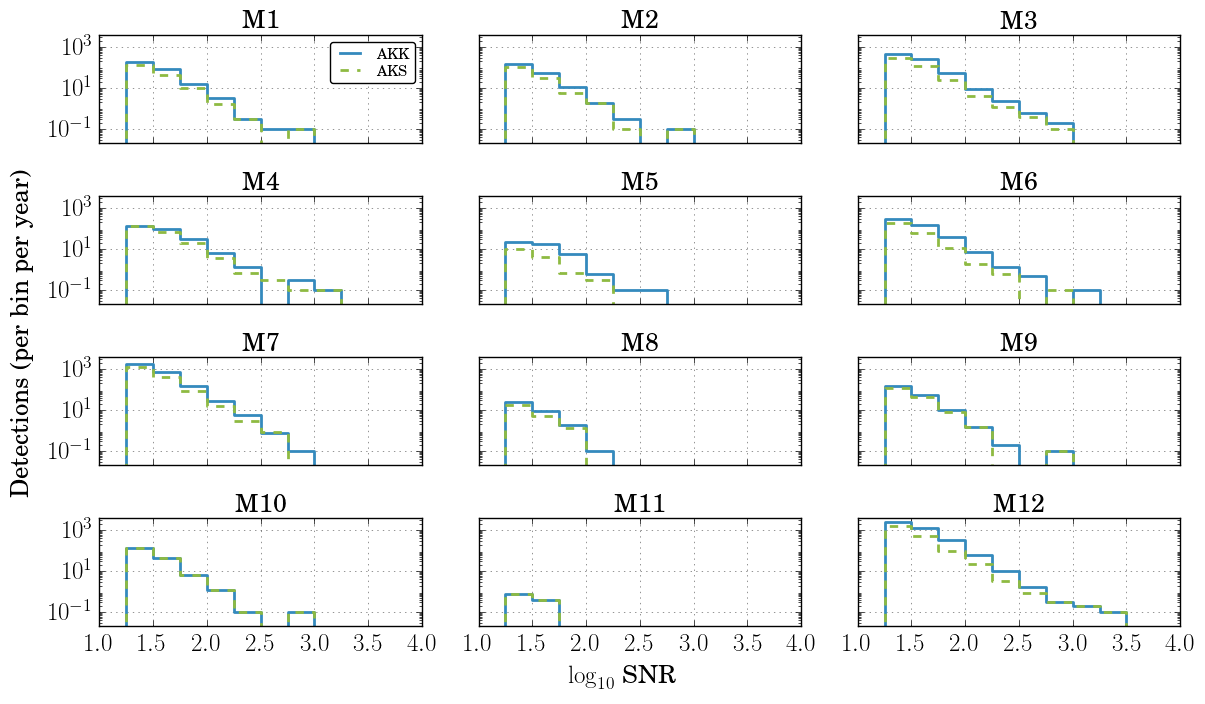}
\caption{SNR distribution for detectable events with AKS and AKK waveforms for all considered models.}
\label{snr}
\end{figure*}

\begin{figure}
\centering
\includegraphics[width=\columnwidth,clip=true,angle=0]{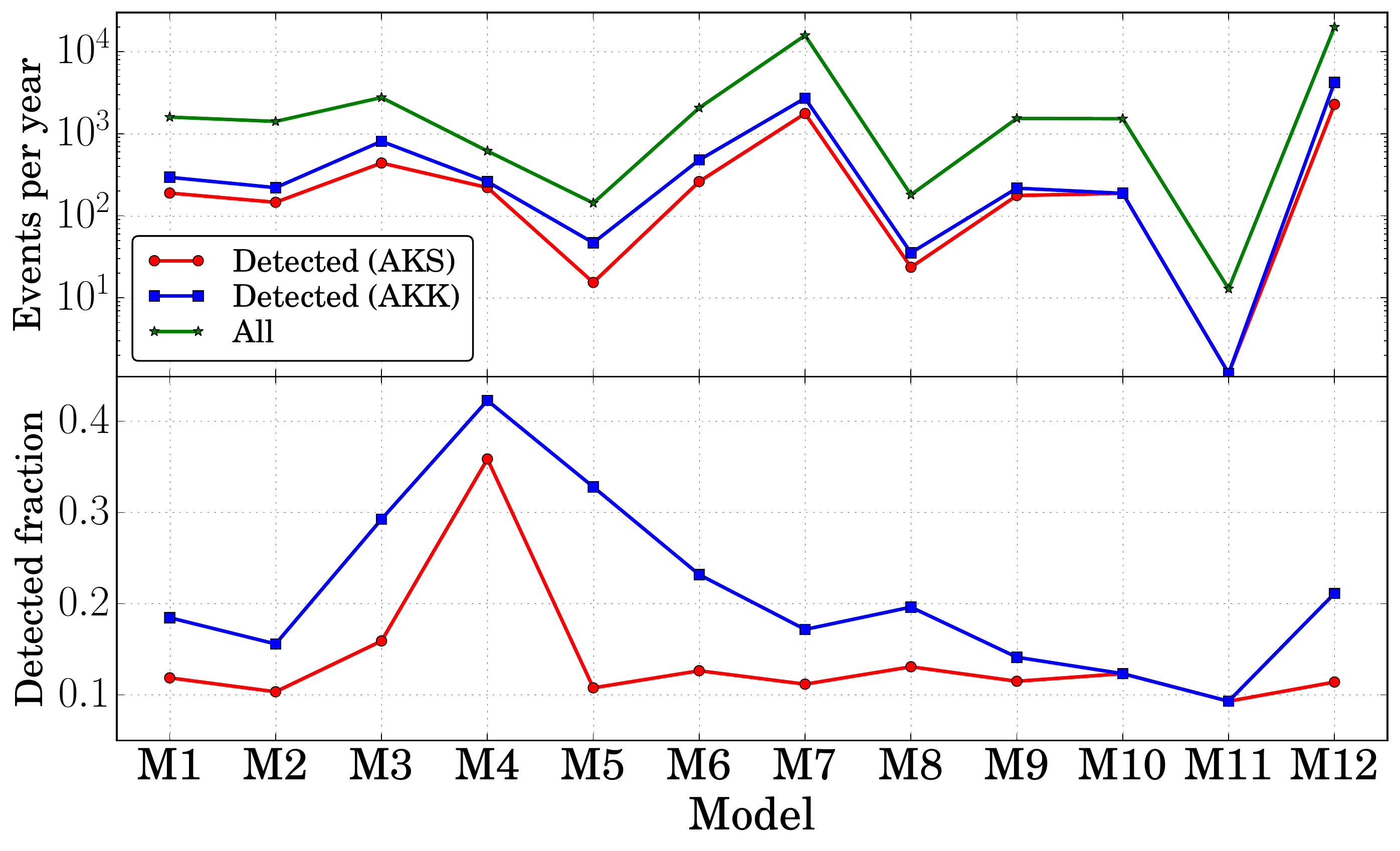}
\caption{Top panel: Event rates (detected and intrinsic) as function of the astrophysical model. Lower panel: Fraction of detection with $\rho>20$ with respect to the total number of EMRIs featuring a central MBH with $10^4\msun<M<10^7\msun$, considering all events at $z<4.5$ ($z<6.5$ for M4).}
\label{rates}
\end{figure}

\begin{figure*}
\centering
\includegraphics[width=\textwidth,clip=true,angle=0]{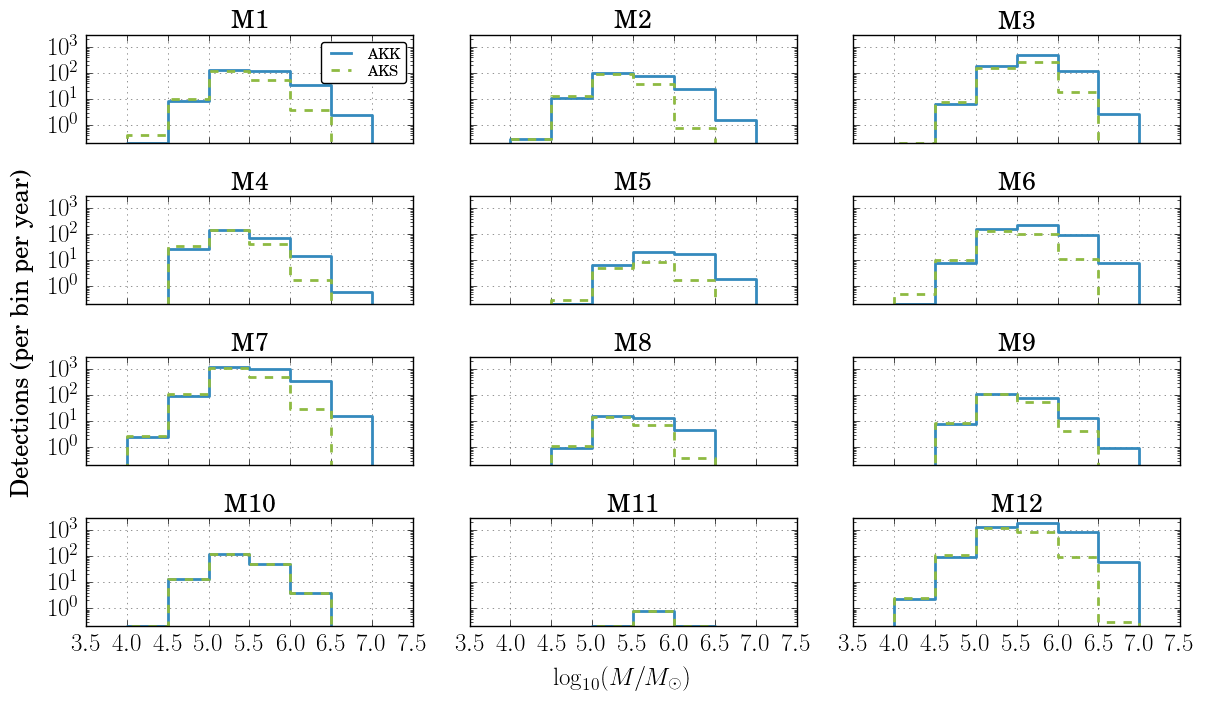}
\caption{MBH mass distribution for detectable events with AKS and AKK waveforms.}
\label{mBins}
\end{figure*}

\begin{figure*}
\centering
\includegraphics[width=\textwidth,clip=true,angle=0]{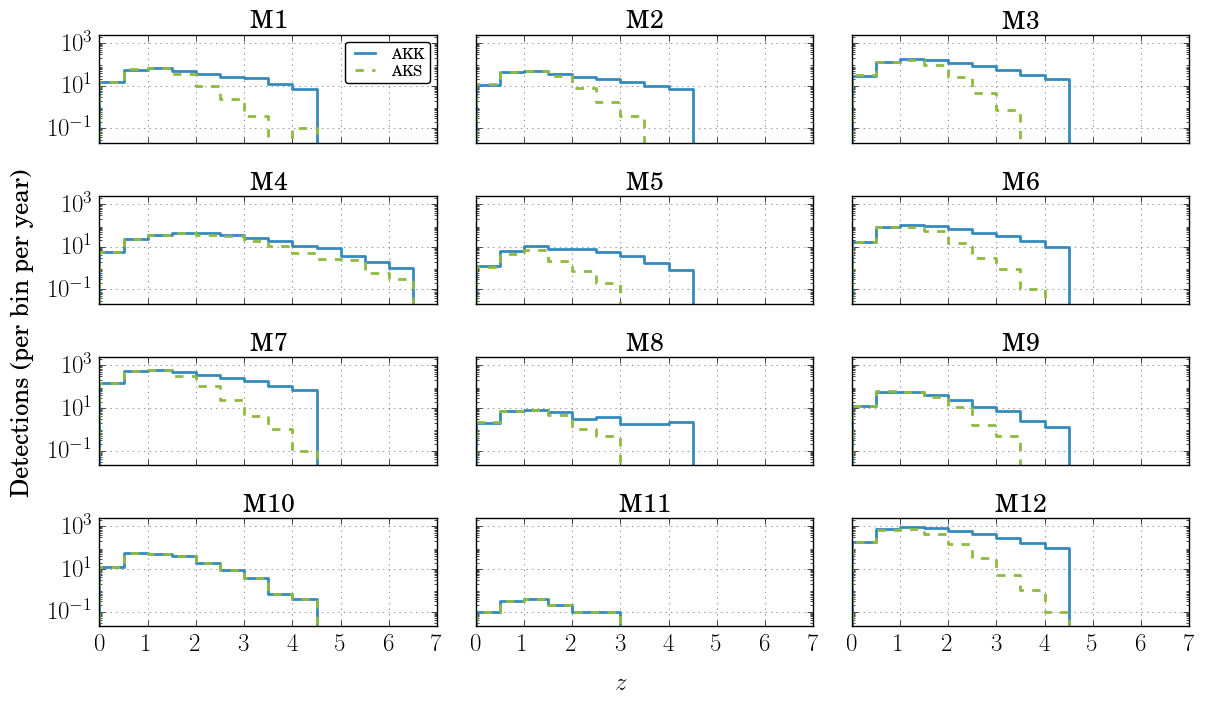}
\caption{Redshift distribution for detectable events with AKS and AKK waveforms. A maximum source redshift of $4.5$ is assumed for all models except M4, where the maximum redshift is $6.5$.}
\label{zBins}
\end{figure*}

\begin{figure*}[htb]
\centering
    \includegraphics[width=\columnwidth]{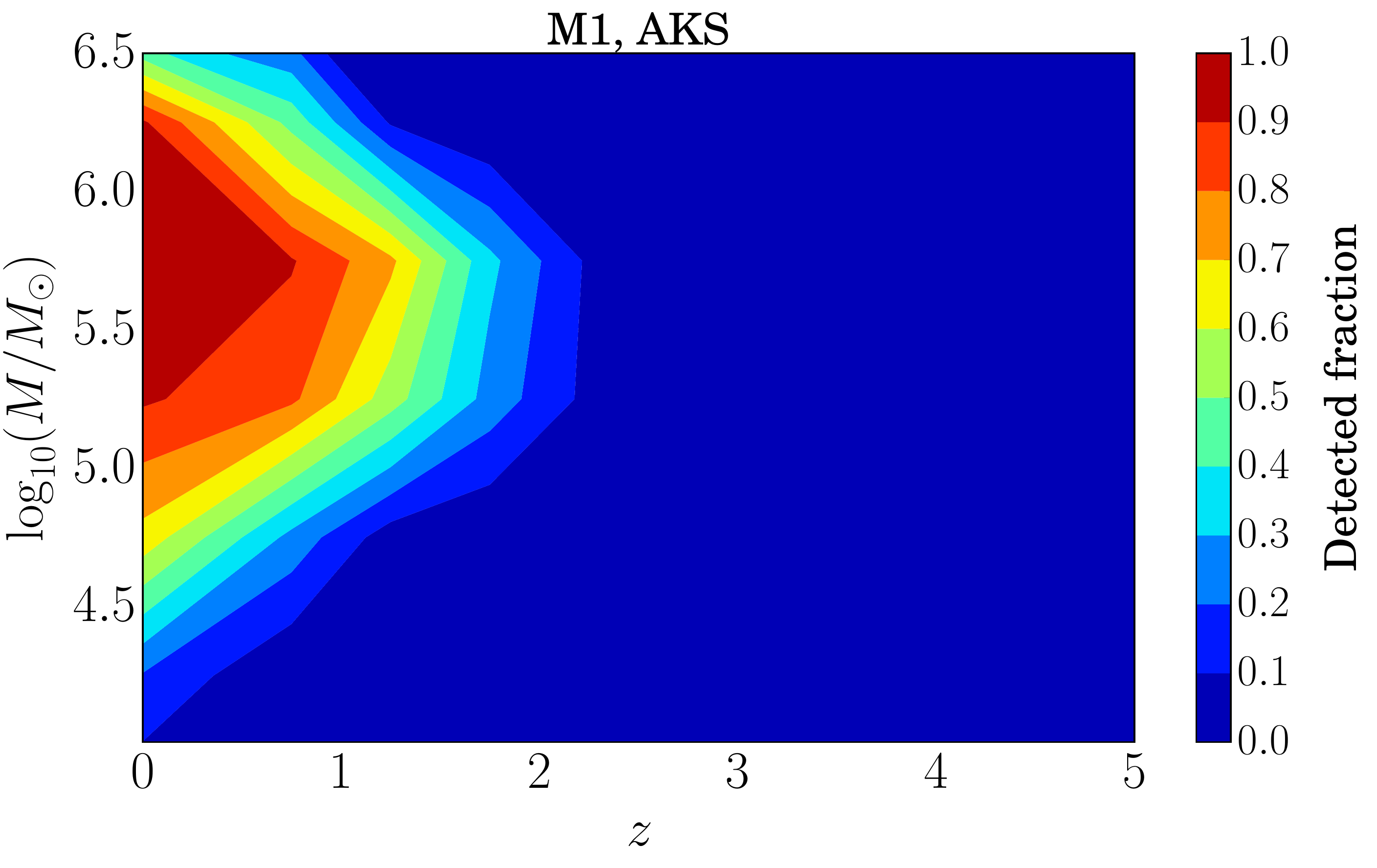} \quad \includegraphics[width=\columnwidth]{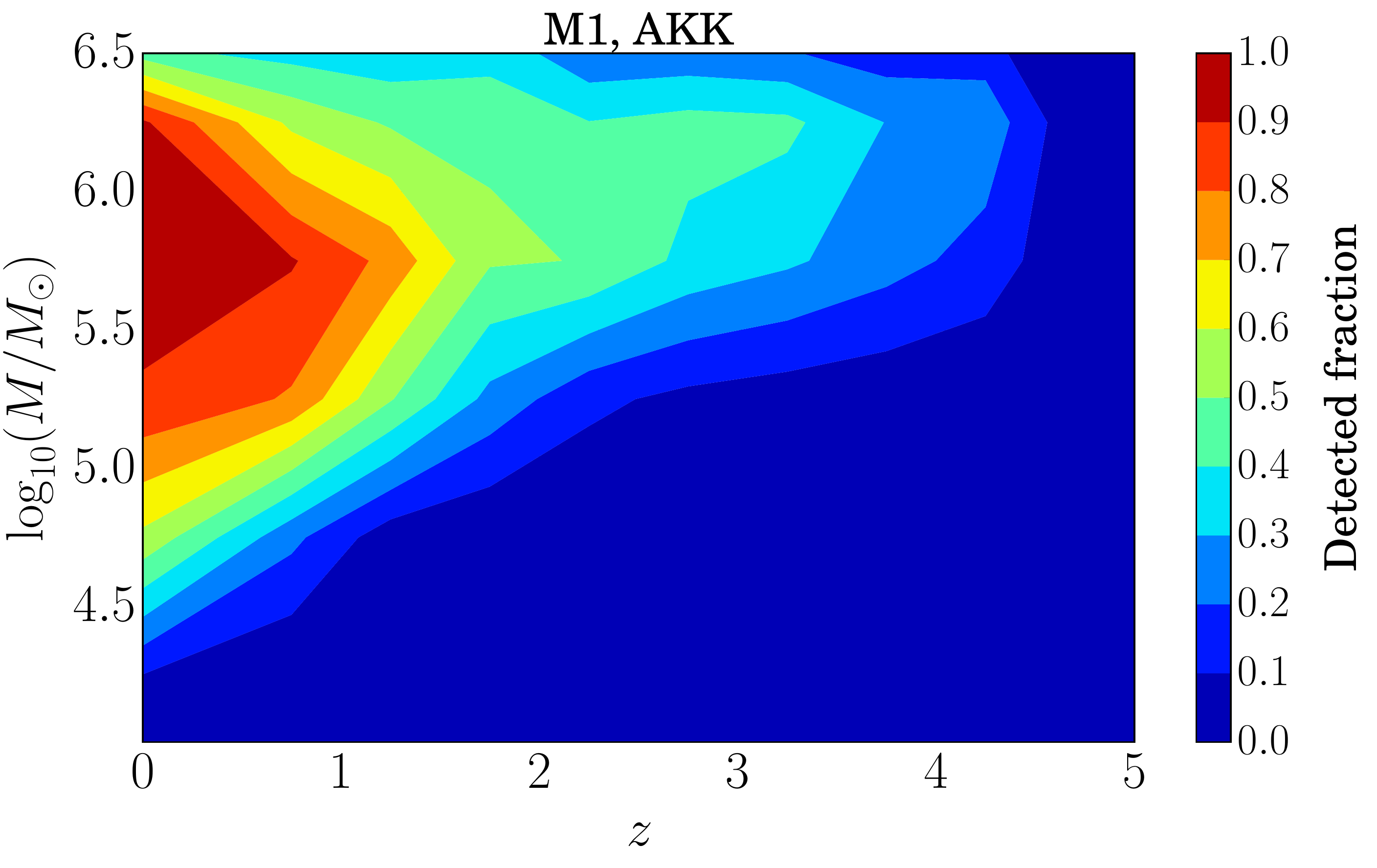}  \\
     \includegraphics[width=\columnwidth]{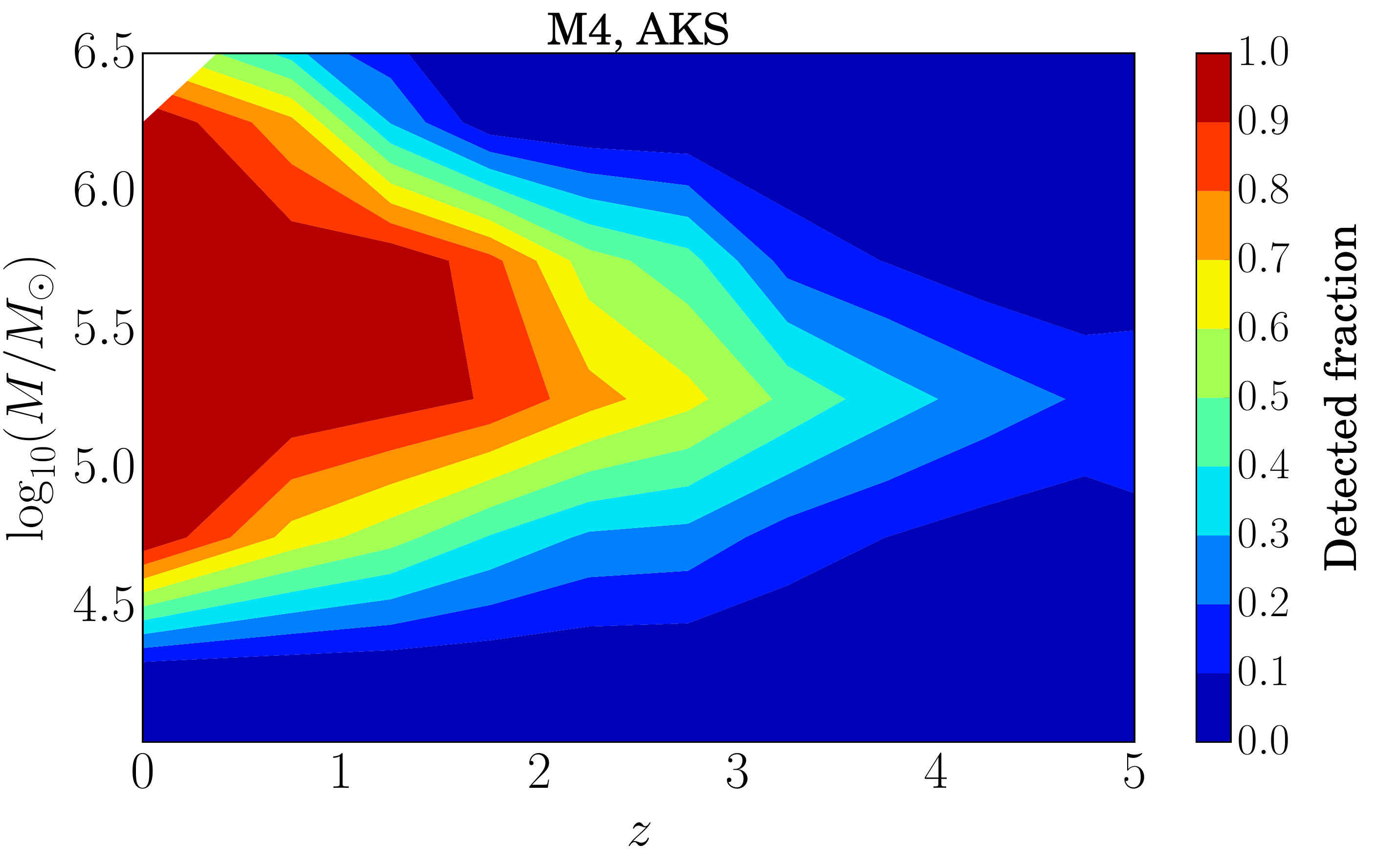} \quad \includegraphics[width=\columnwidth]{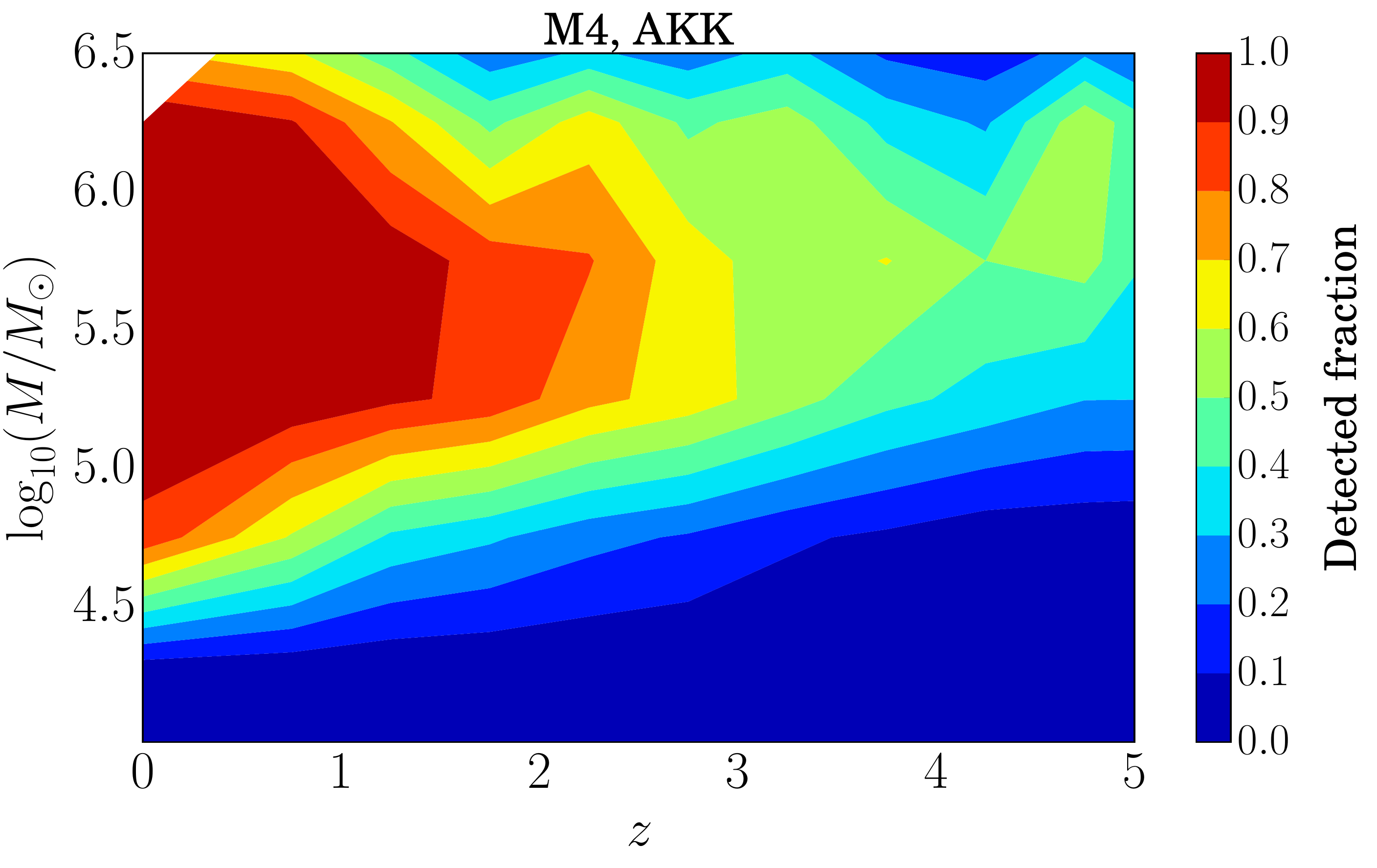}  \\ 
  \caption{Fraction of the intrinsic EMRI population detectable by LISA as a function of source-frame total mass and redshift, for models M1 and M4 and with AKS and AKK waveforms.}
\label{det_frac}
\end{figure*}

With a number of astrophysically motivated EMRI populations and models of the EMRI waveforms in hand, we are all set to investigate the performance of the LISA detector. The AKS and AKK waveforms introduced in Section \ref{sec:analysis} are likely to respectively under- and over-estimate typical EMRI SNRs. In the absence of a more accurate, computationally inexpensive waveform model, we present results for both, with the understanding that they likely bracket the true performance.

\subsection{Detection rates}
To convert from the intrinsic number of EMRIs summarized in Table~\ref{tab1} to the number of LISA detections, 
we must compute the SNR of the GW signal and compare to a detection threshold, which we take to be an SNR of $\varrho = 20$, as mentioned earlier. The SNR calculation depends on the waveform model; we expect the AKK waveforms to produce larger SNRs due to the extrapolation to the Kerr ISCO, resulting in more detectable EMRIs and up to higher redshifts. The SNR distribution of the events above detection threshold is shown in Figure~\ref{snr}. The distribution approximately follows the characteristic $\dd N/\dd \log\rho\propto\rho^{-3}$ behavior of sources uniformly distributed in (Euclidean) volume~\cite{2011CQGra..28l5023S}, with small deviations due to cosmological evolution. As expected AKK waveforms lead to a larger number of detections, which is reflected in the higher normalization of the distribution. For models predicting several hundred sources per year (cf.\ Table~\ref{tab1}), we predict few events in the tail of the distribution, extending to SNR of a few hundred. 

Figure~\ref{rates} shows the number of detectable signals by using the AKS and AKK waveform models (the rates are also reported in the last two columns of Table~\ref{tab1}), and compares these to the total intrinsic rates reported in Table~\ref{tab1}. Based on the Teukolsky horizons shown in Figure~\ref{teuk}, and to save computational time, we considered EMRI populations up to $z=4.5$ for all models with CO mass of $10\msun$, and up to $z=6.5$ when the CO mass is $30\msun$ (M4). We will see below that these maximum redshifts are not sufficient to capture all systems detectable using AKK waveforms. We consider this acceptable since AKK waveform generally overestimate EMRI SNR, and the number of missing events amount to at most a few percent, and thus do not significantly impact our results.
As expected, the rates calculated with the AKK model are generally larger because they produce larger SNRs for spinning MBHs. Models M10 and M11 predict the same detectable rates with AKS and AKK waveforms, since they assume that the MBH spins are zero, in which case AKS and AKK waveforms coincide. When using the AKS model the fraction of detectable events is about $10\%$, independent on the exact features of the model, except for M4 where it increases to around $35\%$. For the AKK waveform, different spin distributions result in different detection fractions, but these still fall between $10\%$ and $20\%$ in most cases. The expected detection rate is therefore roughly proportional to the intrinsic EMRI rate and ranges between $1~\mathrm{yr}^{-1}$ and $2000~\mathrm{yr}^{-1}$ due to severe uncertainties in EMRI astrophysics and dynamics, as discussed in Section \ref{sec:model}.

The fraction of detected events provides clear evidence that EMRI distributions are largely self-similar across the different models, which is confirmed by the (source-frame) mass and redshift distributions of the detected events shown in Figure~\ref{mBins} and Figure~\ref{zBins}. The sharp $z=4.5$ cut-off for the AKK case is due to the maximum redshift of the generated population and not to an intrinsic limitation in the detectability of high redshift sources; the small fraction of the number of missing sources should not significantly impact our results. The most common MBH mass is typically between $10^5 \msun$ and  $10^6 \msun$ in all models. The results based on the AKK waveforms show the detection of more EMRIs into MBHs of larger mass (up to $10^7 \msun$), when MBHs are spinning. This is because for such high mass MBHs a prograde inspiral generates a significant number of waveform cycles between the Schwarzschild ISCO frequency and the final plunge, and these cycles are at frequencies in the most sensitive range for the LISA detector. Thus, the AKS waveforms omit a significant fraction of the SNR for such systems and underestimates their detectability. 
This extra contribution to the SNR also allows sources to be seen to further redshift, as illustrated in Figure~\ref{zBins}.

Taken together, Figures~\ref{mBins} and \ref{zBins} show that EMRI observations will cover MBHs of $3\times 10^4\msun<M<3\times 10^6\msun$ over a redshift range that is broadly peaked at $0.5<z<2$, thus probing a region of the MBH mass--redshift plane that is complementary to both electromagnetic probes of galactic nuclei and LISA observations of MBH binaries. Conventional electromagnetic observations at these low masses out to $z\approx 2$ are extremely challenging, whereas the bulk of LISA MBH binary observations are expected to be at $z>5$, with only few events expected at $z<2$ (cf.~\cite{2011PhRvD..83d4036S}). EMRIs are a unique opportunity to obtain a large sample of confirmed MBHs at relatively low redshift. Figure~\ref{mBins} further highlights that the number of detected EMRIs is sensitive to the minimum mass scale of nuclear MBHs (Alexander and Bar-Or~\cite{2017arXiv170100415A} recently proposed a universal lower limit of about $2\times 10^5\msun$), but in the majority of the investigated models, we predict a few detections at $M>10^6\msun$, which is a relatively safe mass range as it has already been explored by MBH measurements in the local Universe (see, e.g., \cite{2013ApJ...768...76S}).

Examples of LISA's completeness as an EMRI survey are given in Figure \ref{det_frac}, where we plot the fraction of detected sources in the (source-frame) mass--redshift plane for selected models. In the default M1 case, LISA will provide an essentially complete survey in the $10^5 \msun$--$10^6 \msun$ mass range, out to $z\approx 1$, and it is still $50\%$ complete at $z\approx 3$ when AKK waveforms are considered. If inspiralling COs are massive (M4), the survey is complete out to $z\approx 2$ and still $50\%$ complete out beyond $z\approx 4$ for AKK waveforms.

\subsection{Parameter estimation}

\begin{figure*}[htb]
\centering
    \includegraphics[width=\columnwidth]{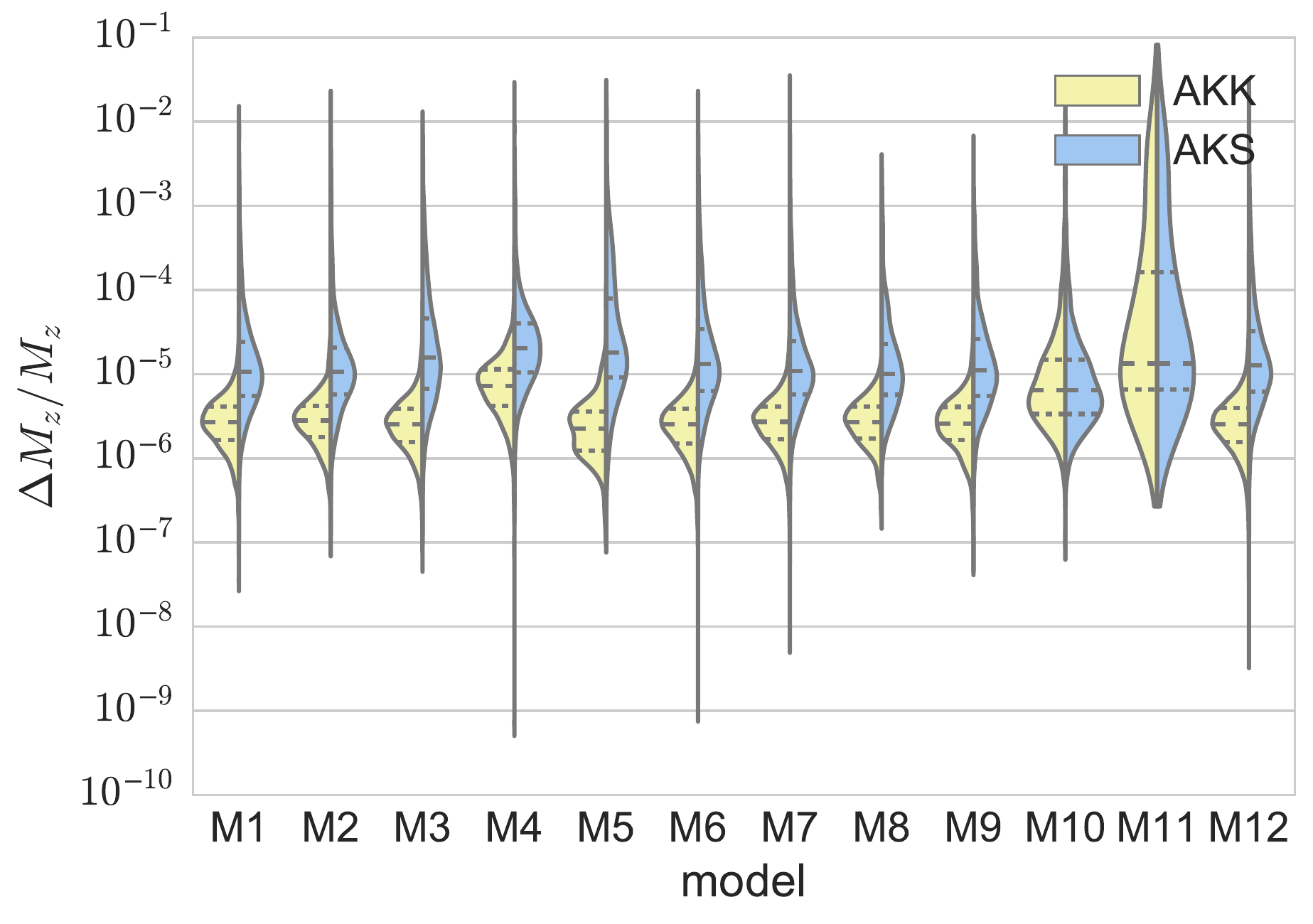} \quad \includegraphics[width=\columnwidth]{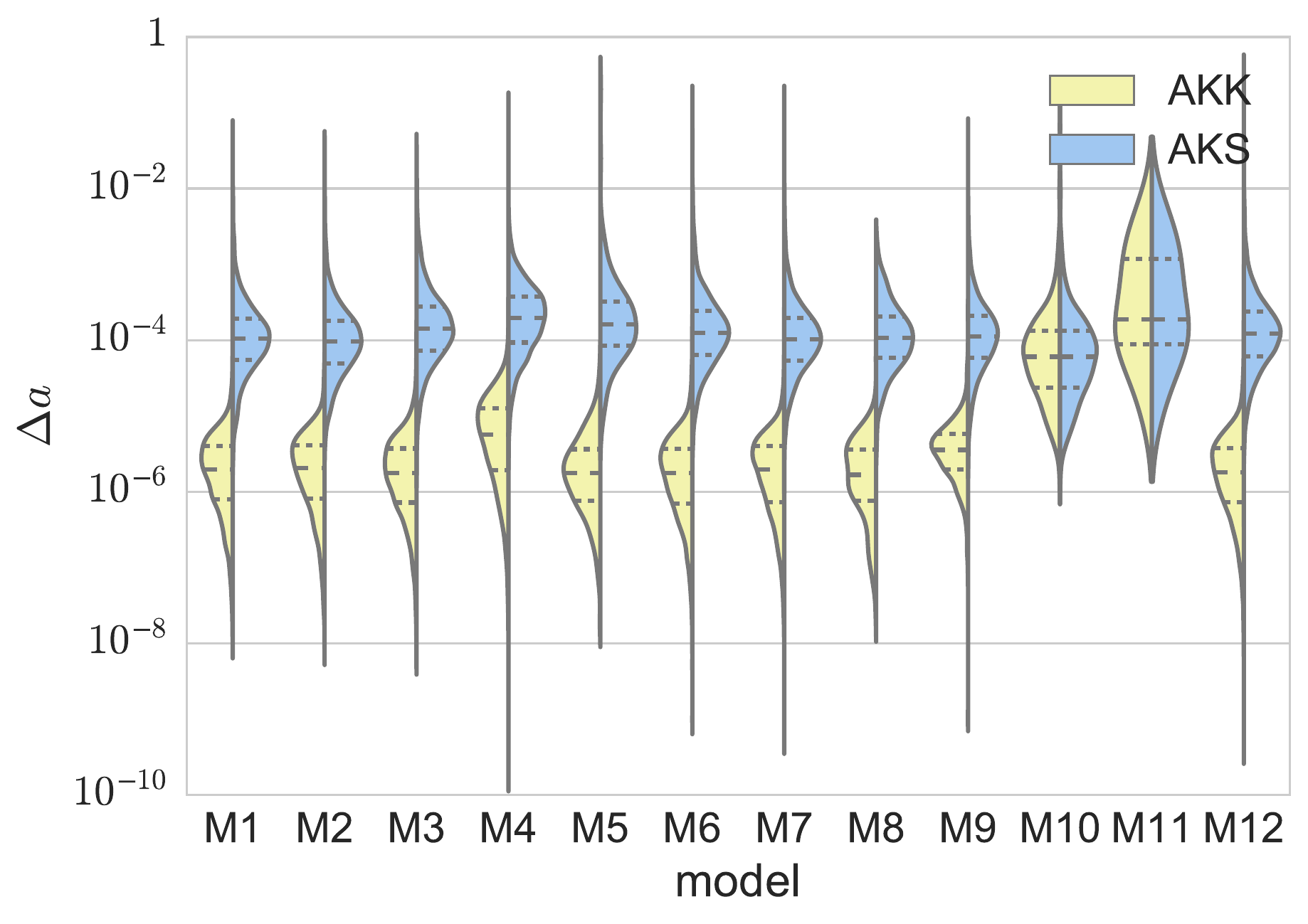} \\
    \includegraphics[width=\columnwidth]{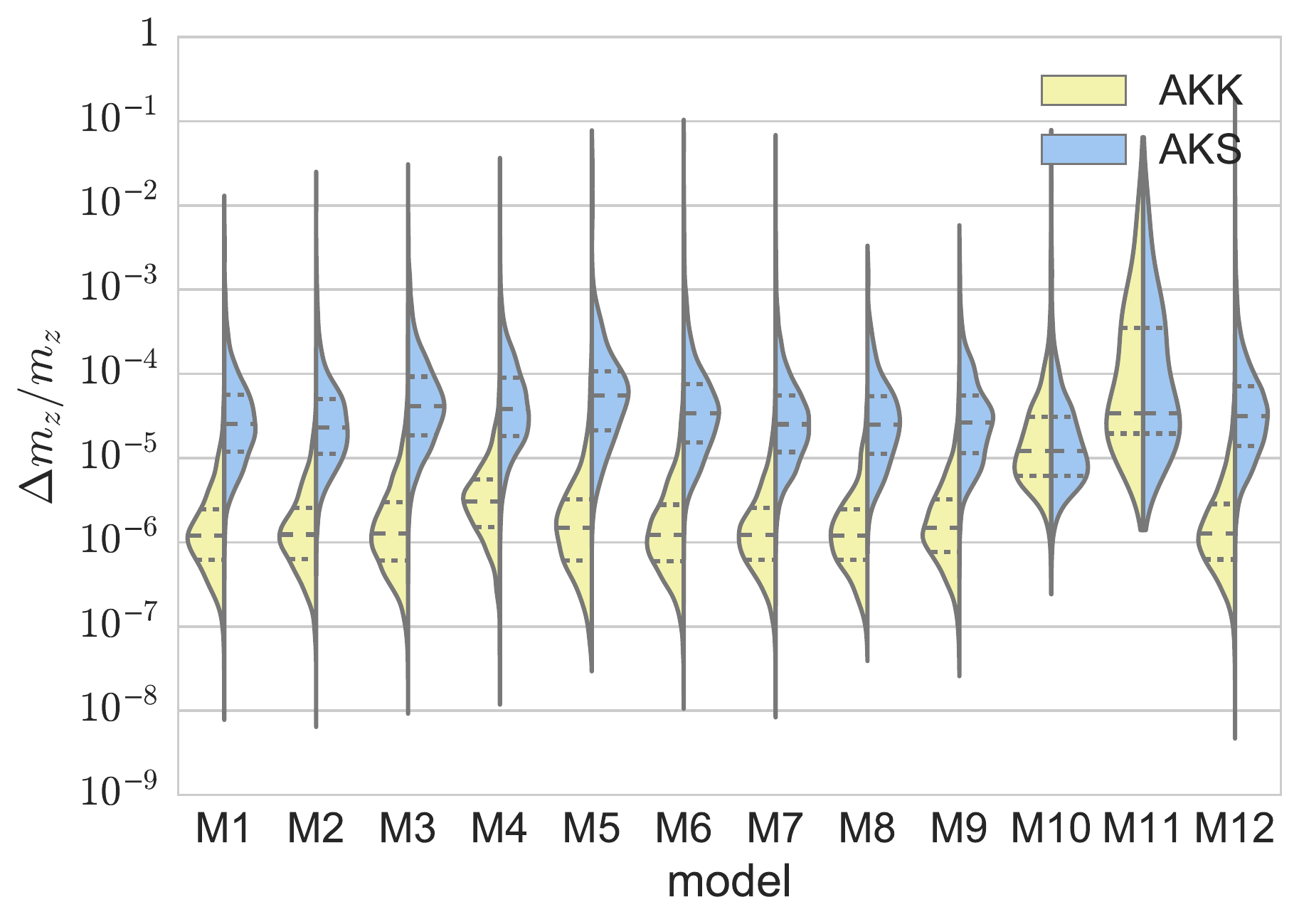} \quad  \includegraphics[width=\columnwidth]{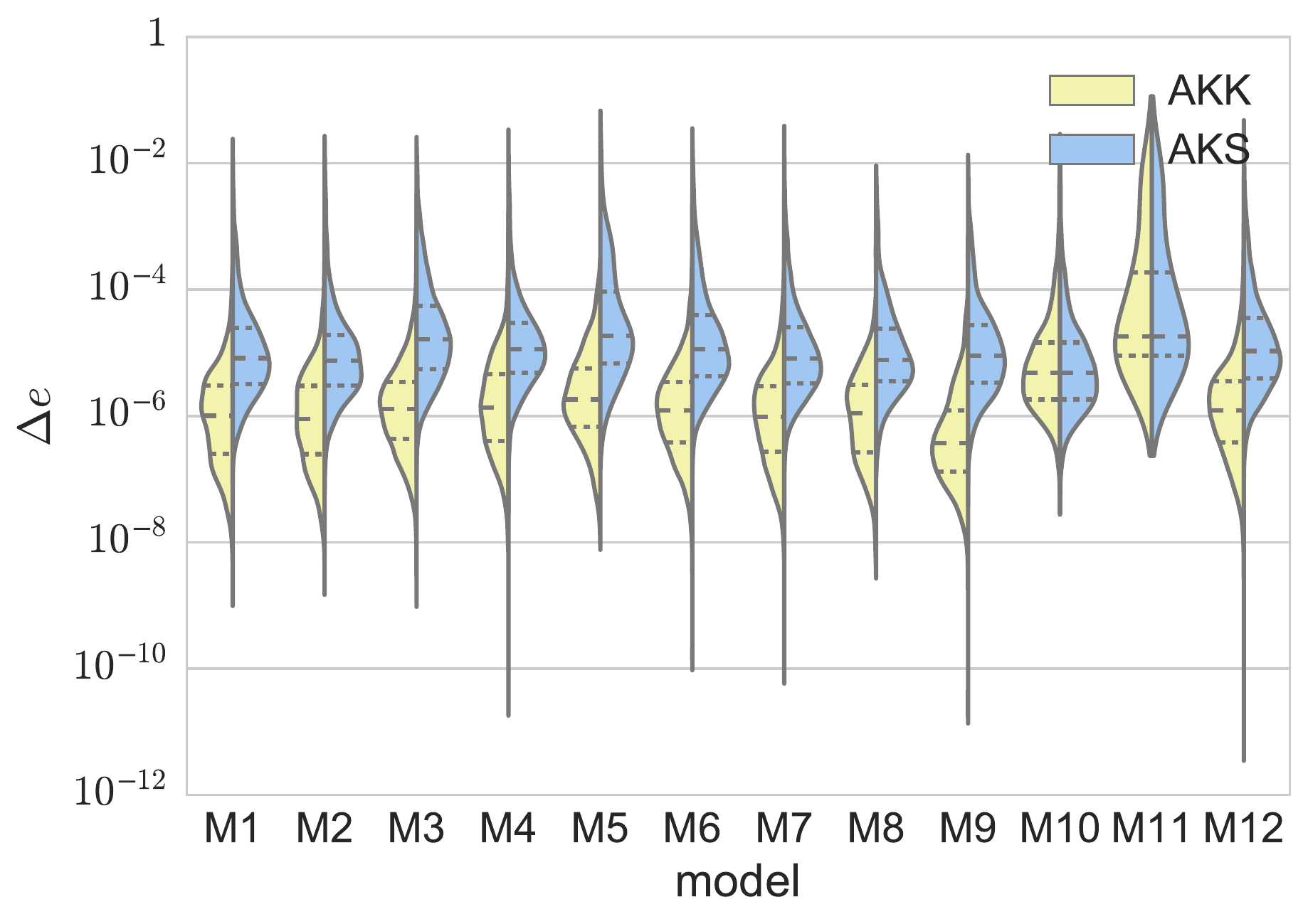}   
    \caption{Distribution over observed EMRIs of the expected statistical errors ($1\sigma$ uncertainties as computed using the Fisher matrix) in the measurement of intrinsic parameters: central MBH redshifted mass (top left), spin (top right), CO mass (bottom left) and eccentricity at plunge (bottom right). The dashed lines mark the first, second and third quartile of the distributions.}
\label{intrinsic}
\end{figure*}

\begin{figure*}[htb]
\centering   
    \includegraphics[width=\columnwidth]{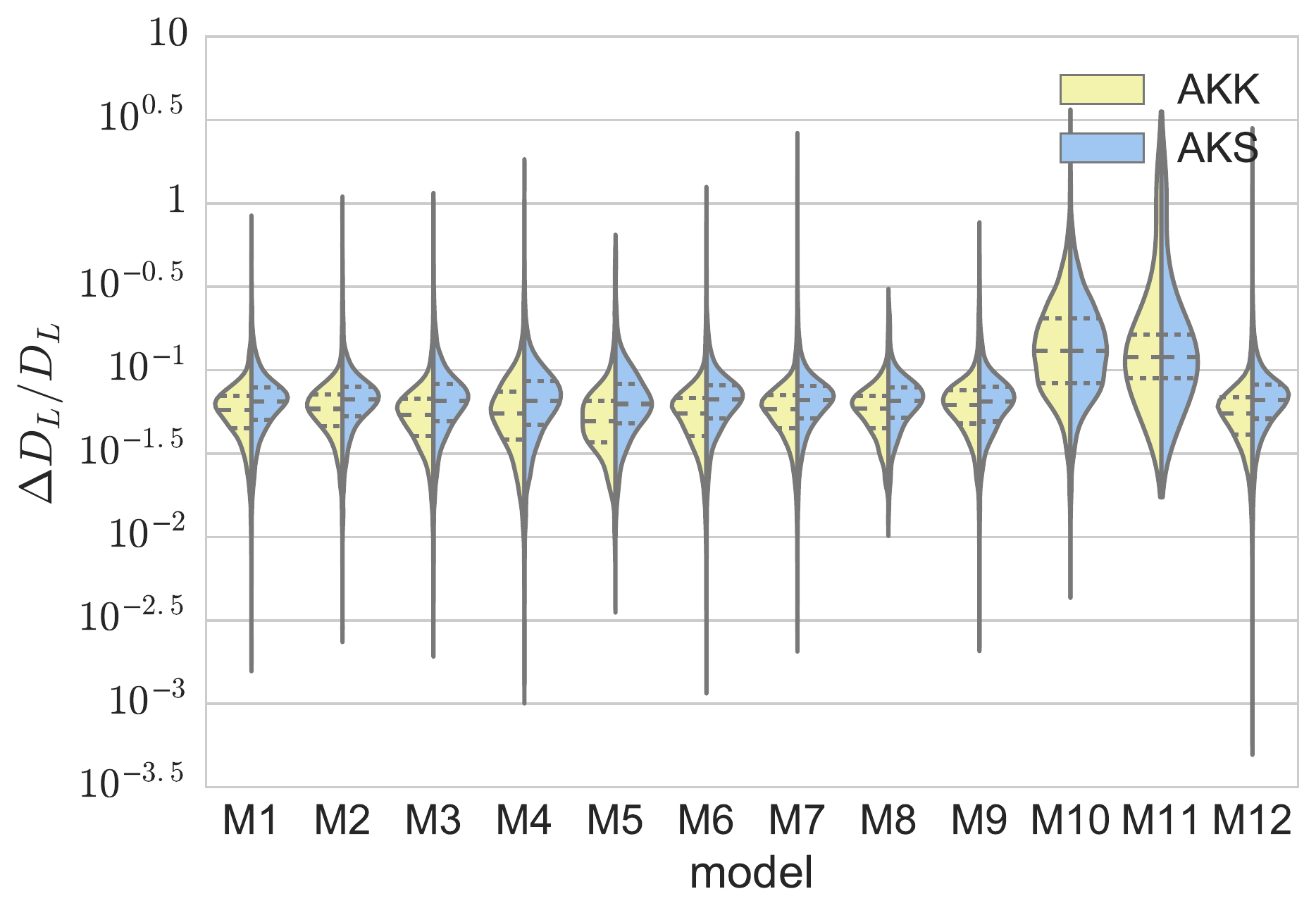} \quad \includegraphics[width=\columnwidth]{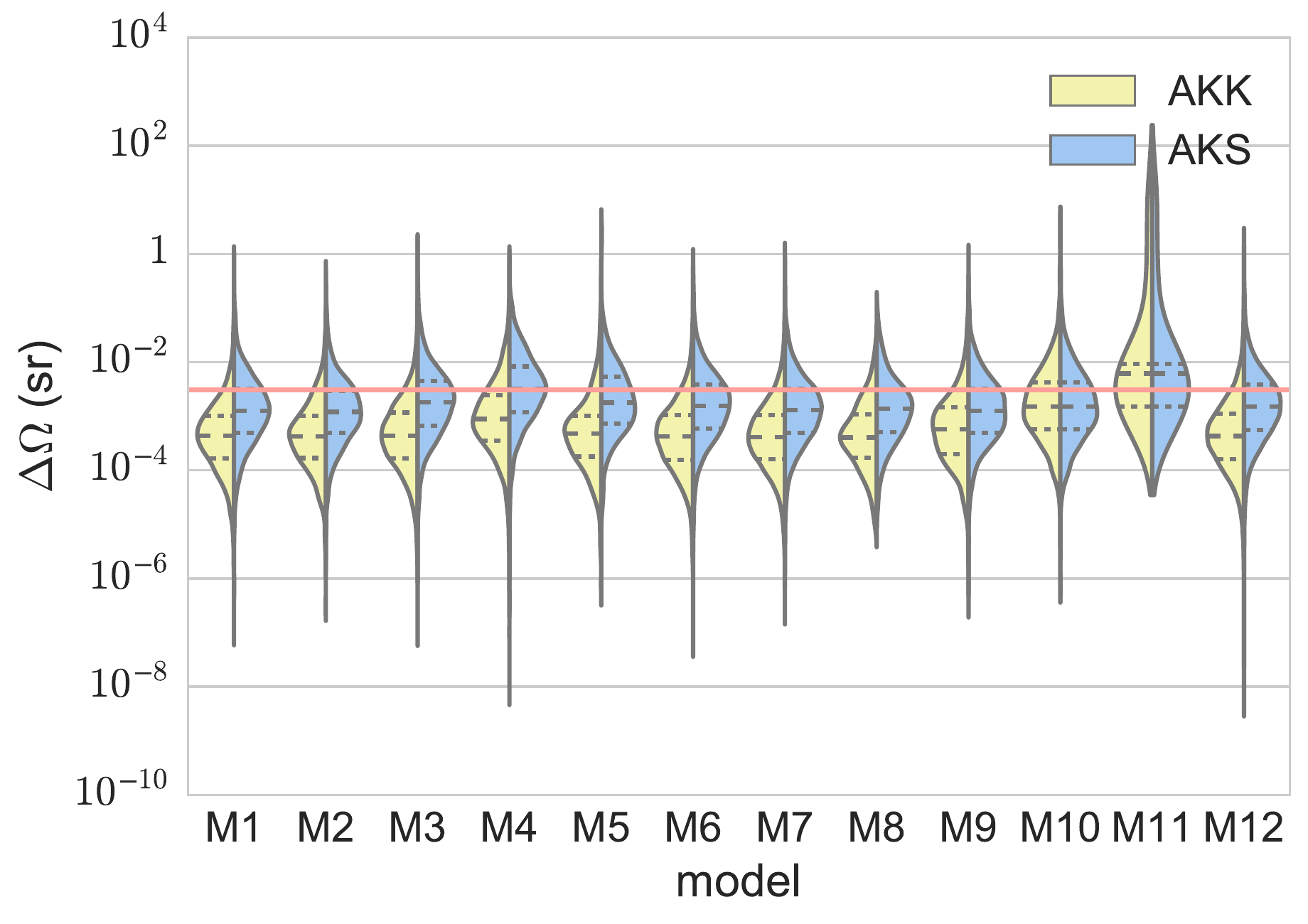}   \\
  \caption{Distribution of the statistical errors in the measurement of EMRI extrinsic parameters: luminosity distance (left panel; $1\sigma$ uncertainty as computed using the Fisher matrix) and sky localization (right panel; the area of an ellipse with probability $1 - \exp(-1)$ of containing the source). The dashed lines mark the first, second and third quartile of the distributions. In the plot for the sky position, a horizontal solid red line marks an error of $10~\mathrm{deg}^2$.}
\label{extrinsic}
\end{figure*}

 \begin{figure}[htb]
\centering   
    \includegraphics[width=\columnwidth]{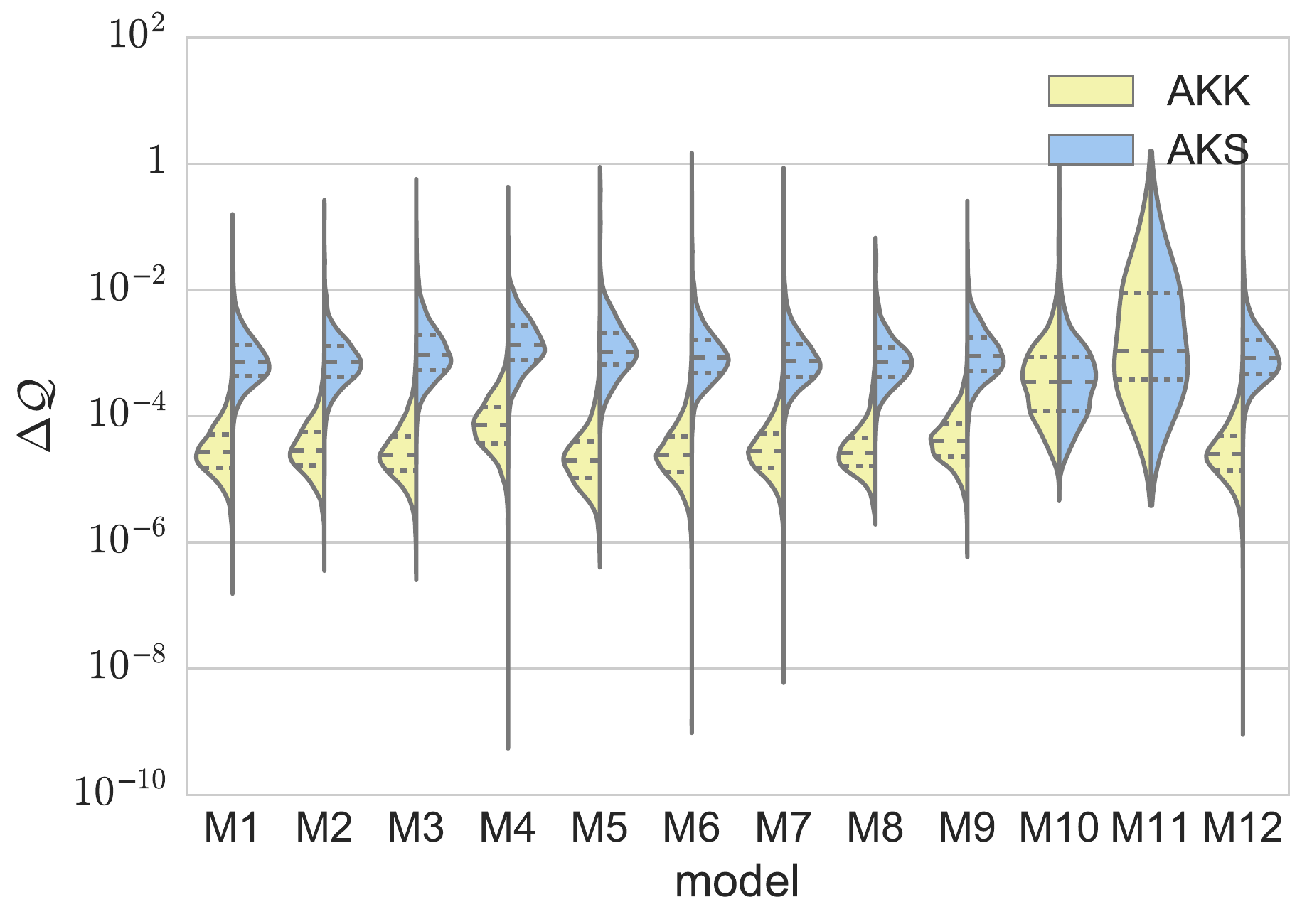}
  \caption{Distribution of the statistical error ($1\sigma$ uncertainties as computed using the Fisher matrix) in measurement of the  deviation of the MBH's quadrupole moment away from the Kerr value. The dashed lines mark the first, second and third quartile of each distribution.}
\label{Kerr}
\end{figure}

Typical EMRIs spend $\mathcal{O}(10^5)$ orbits in the LISA frequency band, and key parameters of the system are encoded in the fine details of the waveform phasing modulation (see, e.g., \cite{2004PhRvD..69h2005B}). 
The redshifted MBH mass $M_z$ sets the characteristic observed frequencies for the EMRI. The rate of inspiral is controlled by the mass ratio, and so gives constraints on the redshifted CO mass $m_z$. 
The MBH spin $a$ also influences the orbital frequencies, and becomes more important as the inspiral gets closer to the MBH; the spin sets the LSO and the transition to plunging.\footnote{The end of the waveform, when the CO plunges into the MBH, also encodes some information which is not captured by these Fisher-matrix estimates~\cite{2014CQGra..31o5005M}. However, the instantaneous SNR in an EMRI is sufficiently low that the plunge is not well resolved, and therefore the inclusion of the plunge should make little difference to parameter estimation.} 
The orbital eccentricity $e$ also affects the orbital frequencies; GW emissions tends to circularize the orbit, so eccentricity is more noticeable earlier in the inspiral.

The large number of cycles completed during the inspiral allow us to obtain exquisite constraints on all the intrinsic parameters, as shown in Figure~\ref{intrinsic}.\footnote{All uncertainties are $1\sigma$ values, except for the sky-localization error $\Delta\Omega$, which is the area of an error ellipse for which there is a probability $\exp(-1)$ of the source being outside of it.}  Even in the conservative AKS case, the median relative error on both redshifted masses is in the range $10^{-4}$--$10^{-5}$ for essentially all models; the spin of the central MBH and the eccentricity at plunge are measured to an absolute precision of about $10^{-4}$ and $10^{-5}$ respectively. Parameter-estimation precisions for the intrinsic parameters are generally better when calculated using AKK waveforms than the AKS waveforms (except for the nonspinning models M10 and M11, for which the AKS and AKK waveforms, and their parameter-estimation errors, coincide). 
This is because of the additional information coming from the late inspiral near the Kerr LSO. 
The difference is most pronounced for the CO mass and MBH spin (for which the measurement improves by a factor of $\approx 30$ on average). 
The difference is less striking for the MBH mass and (especially) the eccentricity: estimation for the latter improves on average by less than a factor of $10$ because it is mostly constrained by the early inspiral. 


Extrinsic parameters such as sky location and distance are primarily determined through the signal amplitude and its modulation as LISA orbits the Sun. These parameters are not strongly dependent upon the GW phase, and hence the large number of cycles completed by an EMRI does not translate to high-precision measurements here. Figure~\ref{extrinsic} shows that the precision of extrinsic parameter measurement is essentially insensitive to the waveform model. The SNR of an individual source may be higher using the AKK waveform, but the overall distribution of SNRs remains largely the same as more quiet signals become detectable~\cite{2011CQGra..28l5023S}, and the typical precision in parameter determination is unaffected. 
On average, the luminosity distance is  measured to $5$--$10 \%$ precision. The luminosity distance is required to convert the observed redshifted masses back to their true source values. Distance uncertainty will therefore be the dominant source of uncertainty in mass measurements.

The sky localization is usually better than $10~\mathrm{deg}^2$. This is the typical field of view of future large optical and radio facilities such as the Large Synoptic Survey Telescope~\cite{2009arXiv0912.0201L} or the Square Kilometre Array~\cite{2009IEEEP..97.1482D}. EMRIs localized to this accuracy can therefore be covered with a single pointing to check for the possible presence of electromagnetic counterparts, which could be associated with the interaction between the CO and an MBH accretion disk~\cite{2008PhRvD..77j4027B,2014PhRvD..89j4059B,2015JPhCS.610a2044B,2011PhRvL.107q1103Y}. Electromagnetic counterparts would be easiest to observe from close by sources, which would also be the loudest, and so the best localized (usually to better than $1~\mathrm{deg}^2$). Identifying a source galaxy from an electromagnetic counterpart would allow for an independent redshift measurement, which would improve the precision of the (source-frame) mass measurements.


Finally, the precise measurements provided by EMRI observations allow us to maps the spacetime of the MBH and check its Kerr nature. 
The multipolar structure of the Kerr metric is completely determined by its mass and spin. According the no-hair theorem, the quadrupole moment is given by $Q_\mathrm{K}=-a^2 M^3$~\cite{1974JMP....15...46H} (see e.g.~\cite{2013LRR....16....7G} for a review of tests of the no-hair theorem with LISA). Since EMRIs are 
expected to probe the multipolar structure of the central MBH spacetime to high 
accuracy~\cite{1995PhRvD..52.5707R,1997PhRvD..56.1845R,1997PhRvD..56.7732R,2006CQGra..23.4167G,2007PhRvD..75d2003B}, 
they will be able to confirm if the quadrupole moment obeys the expected Kerr relation~\cite{2016CQGra..33e4001Y}. In Figure~\ref{Kerr} we show the precision with which possible deviations $\mathcal{Q}$ 
away from the Kerr quadrupole can be constrained. We plot the error on the dimensionless quantity $\mathcal{Q}\equiv (Q-Q_\mathrm{K})/M^3$ (which is independent of the redshifting of masses). 
We do not consider any particular modified theory of gravity: the parameter $\mathcal{Q}$ is just a phenomenological parametrization of hypothetical deviations from the general-relativistic quadrupole moment, and we are interested in determining what level of deviation would be measurable.
As expected, $\mathcal{Q}$ is better constrained by using AKK waveforms, since the effect of a modified quadrupole become important only at small distances from the MBH, i.e.\ in the late inspiral and plunge.

Overall, for all the parameters that we considered, the distributions of the errors are broadly consistent between the different population models. The populations control the number of events, and so are important for considering how much we could learn about the population of MBHs and their host environments, but do not have a significant impact on our ability to extract the parameters for individual EMRIs.

\section{Conclusions}

In this paper we have performed a comprehensive analysis of the performance of the recently proposed LISA mission with regards to the detection and parameter estimation of EMRIs. For the first time
we have attempted to thoroughly investigate the astrophysical uncertainties that affect the calculations of the expected intrinsic EMRI rate. In more detail, we have constructed competing 
astrophysical models for the EMRI rate as a function of cosmic time, accounting
for: the uncertainty on the expected MBH spin magnitude; the disruption of stellar cusps due
to mergers; the MBH growth due to EMRIs and plunges of stellar-mass CO's; and possible viable competing choices for
the MBH mass function, the CO mass, and the correlation between MBH masses and stellar velocity dispersions. Although simple, our models capture the diversity of plausible astrophysical uncertainties. Overall, we find that these astrophysical assumptions produce a variance 
of up to three orders of magnitude in the expected intrinsic EMRI rate. 

For each astrophysical model, we have computed the number of expected detections with the LISA interferometer, as well as the precision with which the source parameters (both intrinsic and extrinsic) can be
recovered. To this purpose, because of computational-time limitations,  we have used two time-inexpensive kludge waveform models~\cite{2004PhRvD..69h2005B} that we expect should bracket the results that would be obtained with more sophisticated Teukolsky or self-force based templates (cf. Fig.~\ref{teuk}). Our main findings are:
\begin{enumerate}
\item Irrespective of the astrophysical model, at least a few EMRIs per year should be detectable by LISA. This number may reach a few thousands per year under the most optimistic 
astrophysical assumptions.
\item Except for the most pessimistic astrophysical models, we predict at least a few events per year should be observable with SNR of several hundreds. 
\item The typical (source-frame) mass and redshift range of detected EMRIs will be $M\sim 10^5$--$10^6 \msun$ and $z\lesssim 2$--$3$, although we may have events with masses an order of magnitude outside of this range or
with larger redshifts (up to $z\sim 4$ and $z\sim 6$ for COs of $10 \msun$ and $30 \msun$, respectively) in all but the most pessimistic astrophysical models.
\item Typical fractional statistical errors with which the intrinsic EMRI parameters (redshifted masses, MBH spin and orbital eccentricity) are expected to be recovered are of the order of $10^{-6}$--$10^{-4}$. Tests of the multipolar structure of the Kerr metric, which only depend upon these mass and spin measurements, can be performed to percent level precision or better. To convert the redshifted masses to the intrinsic source-frame masses requires the luminosity distance, which is typically inferred to $10\%$ precision. Sky localization is usually of the order of a few square degrees. 
It is crucial to model the gravitational waveforms in the late inspiral near the plunge to accurately extract the intrinsic parameters, but this has little impact on the extrinsic parameters.
\end{enumerate} 
These observations could have impact in three distinct areas: astrophysics, cosmology and fundamental physics. 

We have seen that LISA will provide precise measurements of the parameters of individual systems, 
but more information about the astrophysics of these sources will come from studies of populations. 
It was shown in Gair et al.~\cite{2010PhRvD..81j4014G} that the observation of just $10$ EMRIs with 
the classic $5~\mathrm{Gm}$ LISA configuration would be sufficient to measure the slope of the MBH mass 
function in the local Universe to a precision of $\pm0.3$. This is the level to which it is currently 
constrained by electromagnetic observations~\cite{2010PhRvD..81j4014G}. The precision with which LISA can measure EMRI parameters
does not depend strongly on the configuration of the instrument, so this conclusion should carry over to the current analysis. In all the models except the most pessimistic ones, we expect to see many more than $10$ EMRIs, so we would expect to be able to do a high precision measurement of the MBH mass function. One caveat is that what we can measure is the convolution of the MBH mass function with the rate of EMRIs per MBH, not the mass function itself. In Gair et al.~\cite{2010PhRvD..81j4014G} it was assumed that the latter was known, but as we have described here there are many significant uncertainties. It is an open question as to whether these uncertainties can be reduced or at least quantified, or whether LISA observations will be able to decouple them, for instance by using information from the observed MBH mergers. In addition to the MBH mass function, EMRI observations will provide information on the MBH spin distribution, on the properties of the stellar populations in the centers of galaxies and on the relative efficiency of the mechanisms that lead to EMRI formation.

Observations of GW sources provide measurements of the luminosity distance that can be used to measure the expansion history of the Universe~\cite{1986Natur.323..310S}. Individual events do not provide redshifts, but such constraints can be determined statistically. In MacLeod and Hogan~\cite{2008PhRvD..77d3512M} it was shown that if LISA observed $\sim20$ EMRI events at a redshift $z < 0.5$ it would be possible to determine the Hubble constant to better than $1\%$ by using statistical redshifts estimated from galaxy surveys. We find that all but four of our models predict more than $20$ EMRI events at $z < 0.5$.\footnote{Models M4, M5, M8 and M11 do not predict more than $20$ EMRIs at $z < 0.5$. M5, M8 and (in particular) M11 are the pessimistic models which have few EMRI events overall and are almost certainly conservative. M4 is the model with high mass COs, so although this does not predict many EMRIs at low redshift, the reach is much greater, so there is still strong potential for constraints on other cosmological parameters.} However, in MacLeod and Hogan~\cite{2008PhRvD..77d3512M} it was assumed that LISA would determine the luminosity distance and sky location of an EMRI at redshift $z$ to precisions $\Delta(\ln D_\mathrm{L}) < 0.07z$ and $\Delta\Omega < 16 z^2$, which were appropriate for the classic $5~\mathrm{Gm}$ LISA configuration, but are optimistic for the current configuration~\cite{2017arXiv170200786A}. We find that in the models which have $20$ EMRIs at $z < 0.5$, there are at least $5$ that also meet the assumed error constraint. If we used only the events at $z < 0.5$, and with errors smaller than these bounds, we would therefore expect to determine the Hubble constant to at least $\sim2\%$. However, the events with larger errors and events at higher redshift will also contribute to the bound, so we are likely to do better than this, and this should be further explored. In addition, our results show that EMRIs could be detected to much higher redshift than once assumed, which will provide constraints on other cosmological parameters.

The final scientific application of EMRI observations is to tests of fundamental physics. We have already discussed one such application of EMRI observations, the measurement of the quadrupole deviation from the Kerr metric characterized by ${\cal Q}$. Every EMRI will provide a percent or better constraint on that parameter, which is comparable to the expectations for the classic LISA mission configuration. This is no surprise, as the key requirement for a test of fundamental physics is to track the phase of an EMRI over a full inspiral, which has to be done in order to find the EMRI in the data using matched filtering. Thus, any EMRI that is detected will provide a powerful test of fundamental physics, and all of the tests previously discussed in the literature should be possible (see Gair et al.~\cite{2013LRR....16....7G} for a review). Our ability to do this science will not be significantly influenced by the particular astrophysical model, although the models that predict larger rates of EMRI events will more likely lead to the detection of a golden EMRI which is particularly close, has high SNR and, therefore, provides particularly strong constraints.

\acknowledgements
A.~Sesana is supported by the Royal Society. E.~Barausse, E.~Berti and A.~Klein acknowledge support from the H2020-MSCA-RISE-2015 Grant No. StronGrHEP-690904.
This work has made use of the Horizon Cluster, hosted by the Institut d'Astrophysique de Paris. We thank Stephane Rouberol for running smoothly this cluster for us.
E.~Barausse was supported by the APACHE grant (ANR-16-CE31-0001) of the French Agence Nationale de la Recherche.
E.~Berti was supported by NSF Grant No.~PHY-1607130 and by FCT contract
IF/00797/2014/CP1214/CT0012 under the IF2014 Programme.
C.~F.~Sopuerta acknowledges support from contracts ESP2013-47637-P and ESP2015-67234-P (Spanish Ministry of Economy and Competitivity of Spain, MINECO).
C.~P.~L.~Berry is supported by the Science and Technology Facilities Council.
P.~Amaro-Seoane acknowledges support from the Ram{\'o}n y Cajal Programme of the Ministry
of Economy, Industry and Competitiveness of Spain. This work has been partially
supported by the CAS President's International Fellowship Initiative.
This work was supported by the Centre National d'{\'E}tudes Spatiales.

\bibliography{EMRI_paper}

\end{document}